\documentclass[12pt]{article}
\usepackage{amsmath, amsthm,comment}
\usepackage{amsfonts}
\usepackage{amssymb,graphics,psfrag}
\usepackage{array,epsfig,stmaryrd,graphicx}
\usepackage{hyperref}
\usepackage{datetime,color}
\usepackage[export]{adjustbox}

\newcommand{\beq}{\begin{equation}}
\newcommand{\eeq}{\end{equation}}
\newcommand{\beqa}{\begin{eqnarray}}
\newcommand{\eeqa}{\end{eqnarray}}
\newcommand{\beqar}{\begin{eqnarray*}}
\newcommand{\eeqar}{\end{eqnarray*}}

\begin{document}
\baselineskip 18pt%
\begin{titlepage}
\vspace*{1mm}%
\hfill
\vbox{

    \halign{#\hfil         \cr
   %       hep-th/yymmnnn\cr
      %   CERN-PH-TH/2013-xyz\cr
        % IPM/P-2010/003  \cr
         %  CPHT RR-xxx .yyzz \cr
           } % end of \halign
      }  % end of \vbox
\vspace*{10mm}
\vspace*{12mm}%

\center{ {\bf \Large  Estimation of Critical Collapse Solutions to Black Holes with Nonlinear Statistical Models
%Smoothers

}}\vspace*{3mm} \centerline{{\Large {\bf  }}}
\vspace*{5mm}
\begin{center}
{ Ehsan Hatefi$^{\dagger,\star}$\footnote{E-mails: ehsan.hatefi@uah.es, ehsanhatefi@gmail.com, ahatefi@mun.ca} and  Armin Hatefi$^{\ddagger}$}

\vspace*{0.8cm}{ { \footnotesize
$^{\dagger}$ GRAM Research Group, Department of Signal Theory and Communications,
University of Alcala, Alcala de Henares, 28805, Spain.\\

$^{\star}$ Scuola Normale Superiore and I.N.F.N,
Piazza dei Cavalieri 7, 56126, Pisa, Italy.\\
$^{\ddagger}$ Department of Mathematics and Statistics, Memorial University of Newfoundland, St John’s, NL, Canada. }}
\vspace*{1.3cm}
\end{center}
\begin{center}{\bf Abstract}\end{center}
\begin{quote}

The self-similar gravitational collapse solutions to the Einstein-axion-dilaton system have already been found out. 
Those solutions become invariants after combining the spacetime dilation with the transformations of internal SL(2, R). 
We apply nonlinear statistical models to estimate the functions that appear in the physics of Black Holes of the 
axion-dilaton system in four dimensions. These statistical models include parametric polynomial regression, nonparametric kernel 
regression and semi-parametric local polynomial regression models. Through various numerical studies, we reached accurate numerical
 and closed-form continuously differentiable estimates for the functions appearing in the metric and equations of motion.
\end{quote}
\end{titlepage}%

 %%%%%%%%%%%%%%%%%%%%%%%%%%%%%%%%%%%%%%%%%%%%%%%%%%%%%%%%%%%%%%%%%%%%%%%%%% %%%%%%%%%%%%%%%%%%%%%%%%%%%%%%%%%%%%%%%%%%%%%%%%%%%%%%%%%%%%%%%%%%%%%%%%%%

\section{Introduction}

As the end state of gravitational collapse,  black holes  are defined by their mass, angular momentum as well as their charge. 
M. Choptuik \cite{Chop} explored the so called critical phenomena in gravitational collapse as well 
as Choptuik scaling. He made a breakthrough in the subject of numerical relativity. Indeed, Choptuik scaling 
(\cite{Chop} and \cite{Gundlach:2002sx}) is a property that occurs in various systems  which experience 
gravitational collapse. He discovered that there might be a fourth universal quantity that establishes the critical 
collapse. Choptuik followed the study of the spherically symmetric collapse of scalar field and explored a critical 
behaviour which demonstrates the discrete spacetime self-similarity. Through taking the amplitude of the scalar field $p$,
 he derived a critical value  $p_\text{crit}$ where black hole forms as $p$ exceeds $p_\text{crit}$.
 Also, as $p$ goes beyond the threshold, the mass of the black hole $M_\text{bh}$ illustrates the scaling law
\begin{equation}
M_\text{bh}(p) \propto (p-p_\text{crit})^\gamma\,,
\end{equation}
where  the Choptuik exponent was found to be $\gamma\simeq 0.37$~\cite{Chop} in four dimension and for a real scalar field. Various numerical computations with different matter content have also been discovered \cite{Birukou:2002kk,Husain:2002nk,Sorkin:2005vz,Bland:2005vu,Rocha:2018lmv}.

Motivated by String Theory, the axion-dilaton system can also experience the same gravitational collapse process. The study of Choptuik  phenomenon in the axion-dilaton system was initiated in \cite{HE,Hamade:1995jx,Eardley:1995ns}.  
The AdS/CFT correspondence \cite{Maldacena:1997re} is viewed as the first motivation to investigate critical collapse solutions especially for the axion-dilaton system. The AdS/CFT correspondence correlates the critical exponent and the imaginary part of quasi normal modes as well as the dual conformal field theory \cite{Birmingham:2001hc}. 
%The first motivation for studying critical collapse solutions especially for the axion-dilaton system is the AdS/CFT correspondence \cite{Maldacena:1997re}, 
%The AdS/CFT correspondence correlates the critical exponent and the imaginary part of quasi normal modes as well as the dual conformal field theory \cite{Birmingham:2001hc}. 
The second motivation relies on the holographic description of black hole formation \cite{AlvarezGaume:2006dw}, particularly in the physics of black holes and their implications \cite{Hatefi:2012bp}. From the IIB string theory point of view, we look for the gravitational collapse for the special spaces that could asymptotically approach $AdS_5 \times S^5$. The matter field in the IIB string theory arises from the self-dual 5-form field strength and the axion-dilaton configuration.
In a recent research \cite{ours,Antonelli:2019dqv},
 the whole families of Continuous Self-Similar (CSS) solutions of the Einstein-axion-dilaton system  were explored for all the three conjugacy classes of  SL(2,R).
Some remarks about critical exponents and higher dimensional solutions have been made in \cite{Hatefi:2020jdr} and \cite{Hatefi:2020gis}.
For more details about the other systems experiencing gravitational collapse, readers are referred  to  \cite{Hamade:1995ce,AE, EC,KHA, MA}.

To our best knowledge, there is no research article in the literature investigated the properties 
 of nonlinear statistical models to estimate the critical collapse functions in Einstein-axion-dilaton.
In this paper, for the first time, we utilize parametric polynomial regression, non-parametric kernel regression and semi-parametric local polynomial regression models  to develop and a closed form and continuously differentiable  functional forms of the critical collapse functions.

This article is organized as follows. We decribe the axion-dilaton system and its different Continuous Self-Similar ansatz\"e in Section \ref{sec:con}.  
 The initial conditions and Properties of the critical solutions for all three conjugacy classes are discussed in Sections  \ref{sec:ini} and \ref{sec:prop}, respectively. The nonlinear statistical estimation methods are then discussed  in Section \ref{sec:stat}. The performance of proposed statistical models are finally investigated in Section \ref{sec:num}. Concluding remarks are presented in Section \ref{sec:sum}.

\section{Axion-dilaton Configuration}\label{sec:con}

One can combine the axion $a$ and dilaton $\phi$ field into a single complex field
$\tau \equiv a + i e^{- \phi}$, and its coupling to four dimensional gravity is given by
\begin{equation}
S = \int d^4 x \sqrt{-g} \left( R - {1 \over 2} { \partial_a \tau
\partial^a \bar{\tau} \over (\mathop{\rm Im}\tau)^2} \right) \; .
\label{eaction}\end{equation}
where $R$ is the scalar curvature.  The above action describes the effective action of type II string theory \cite{Sen:1994fa,Schwarz:1994xn}. This action respects SL(2,R) symmetry, which means that if we consider the following 
 \begin{equation} \label{eq:sltr}
     \tau \rightarrow  \tau \equiv \frac{a \tau +b}{c \tau + d} \; ,
 \end{equation}
the action remains invariant  where $a$, $b$, $c$, $d$ are real parameters satisfying $ad-bc = 1$.
The equations of motion can be read as follows
\begin{eqnarray}
\label{eoms}
R_{ab} - {1 \over 4 (\mathop{\rm Im}\tau)^2} ( \partial_a \tau \partial_b
\bar{\tau} + \partial_a \bar{\tau} \partial_b \tau) & = & 0 \end{eqnarray}
\begin{eqnarray}
\label{eoms14}
\nabla^a \nabla_a \tau + { i \nabla^a \tau \nabla_a \tau \over
\mathop{\rm Im}\tau} = 0 .
\end{eqnarray}
We have looked for critical solutions by dealing with
spherical symmetry and continuous self-similarity. Following  \cite{HE,Hamade:1995jx,Eardley:1995ns}, one can choose the metric as
\begin{equation}
	ds^2 = \left(1+u(t,r)\right)\left(- b(t,r)^2dt^2 + dr^2\right)
			+ r^2d\Omega^2 \;.
\label{metric1}
\end{equation}

We might consider a scale invariant variable as $z \equiv - r/t$ and hence the Continuous Self-Similarity of the metric actually means that all 
functions $u(t,r),b(t,r)$ can be expressed just in terms of $z$, that is, $b(t,r) = b(z), u(t,r) = u(z)$.

This continuous self-similarity condition for $\tau$  was described in detail in \cite{AlvarezGaume:2011rk}. 
The axion-dilaton system does have a global $SL(2,R)$-symmetry which is broken to a $SL(2,Z)$ by taking into account the
non-perturbative phenomena in type II string theory. 
 If we take the quantum effects, SL(2,R) symmetry reduces to SL(2,Z) for which it is believed to be non-perturbative symmetry of String Theory~\cite{gsw,JOE,Font:1990gx}.
Therefore, one might compensate the action 
% ($\xi = t \frac{\partial}{\partial t} + r \frac{\partial}{\partial r}$)  
by means of an $SL(2,R)$-transformation, that is  $\tau(t,z)$ must respect the following equation
\begin{equation}
t\,{\partial\over \partial t}\,\tau(t,z)\,=\,\alpha_0\,+\,\alpha_1\,\tau\,+\,\alpha_2\,\tau^2
\end{equation}
with $\alpha_{0,1,2}$ real numbers. The above equation
has two roots  that are related to compensating the scaling transformation.  Having set that, we find three different ansatz\"e, which are related to the fact that the chosen $SL(2,R)$-transformation is either an
elliptic, hyperbolic or parabolic transformation. The elliptic ansatz  is defined as \begin{equation}
 \tau(t,r)	=  i { 1 - (-t)^{i \omega} f(z) \over 1 + (-t)^{i
\omega} f(z)} ,
\label{tauansatz}
\end{equation}
where $\omega$ is a real constant that will be known by the regularity conditions for the critical solution.
On the other hand, for the hyperbolic case, $\tau(t,r)$ is given by
\begin{equation}
\tau(t,r) = \;\frac{1- (-t)^{\omega} f(z)}{1+ (-t)^{\omega} f(z)},
\label{tau2ansatz}
\end{equation}
Eventually the parabolic ansatz is illustrated by $\tau(t,r) = f(z)+\omega \log(-t)$.
%where one can show that the the ansatz $\tau(t,r)= (-t)^\omega f(z)$ also leads to the same equations of motion for hyperbolic case,
Note that, the function $f(z)$ needs to satisfy $|f(z)| < 1$ for the elliptic case, whereas $Im f(z)>0$ for hyperbolic and parabolic cases.

\section{Equations of motion and initial conditions} \label{sec:ini}

 In this section, we first study the equations of motion and then explain the properties of solutions. Replacing CSS into the equations of motion, we derive a system of differential equations just for $b(z)$, $f(z)$. Using Einstein equations for angular variables, one can express $u(z),b(z)$  just in terms of $f(z)$, which means that 
\begin{equation}
u(z)\,=\,{z\, b'(z)\over \,b(z)}.
\end{equation}

Hence $u(z)$ and its derivatives can be eliminated from equations of motion. The other equations of motion involve $b(z), f(z)$. Hence we are left with various ordinary differential equations (ODEs) 
\begin{align}\label{eq:unperturbedbp}
    b'(z) & = B(b(z),f(z),f'(z))\,, \\
    f''(z) & = F(b(z),f(z),f'(z))\,. \label{eq:unperturbedfpp} \end{align}
 Since in this paper, we are looking for estimation of the function of  $b(z)$ and real and Imaginary part of $f(z)$ for elliptic and hyperbolic cases in 4 dimension, we just generate those equations as follows. Indeed the equations of motion are derived in  \cite{AlvarezGaume:2011rk}.  The equations for the elliptic case are
\begin{eqnarray}
0 & = & b' + { z(b^2 - z^2) \over b (-1 + |f|^2)^2} f' \bar{f}' - {
i \omega (b^2 - z^2) \over b (-1 + |f|^2)^2} (f \bar{f}' - \bar{f} f')
- {\omega^2 z |f|^2 \over b (-1 + |f|^2)^2}, \nonumber\\
0 & = & f''
     - {z (b^2 + z^2) \over b^2 (-1 + |f|^2)^2} f'^2 \bar{f}'
     + {2 \over (1 - |f|^2)} \left(1
       - {i \omega (b^2 + z^2) \over 2 b^2 (1 - |f|^2)} \right) \bar{f} f'^2 \nonumber \\&&
     + {i \omega (b^2 + 2 z^2) \over b^2 (-1 + |f|^2)^2} f f'
\bar{f}' 
  + {2 \over z} \left(1+ {i \omega z^2 (1 + |f|^2) \over (b^2 - z^2)
(1 - |f|^2)}\right.\nonumber \\&& 
+ \left.{\omega^2 z^4 |f|^2 \over b^2 (b^2 - z^2) (1 -
|f|^2)^2}\right) f'+ {\omega^2 z \over b^2 (-1 +|f|^2)^2} f^2
\bar{f}' + \nonumber \\&&
{2i \omega \over (b^2 - z^2)} \left(\frac{1}{2} - {i \omega (1 + |f|^2)
\over 2(1 - |f|^2)}\right.
- \left.{\omega^2 z^2 |f|^2 \over 2b^2 (-1 + |f|^2)^2}
\right) f.
\label{1fzeom321}
\end{eqnarray}
 Using time scaling one can set $b(t,0)=1$. In the elliptic case,
by writing $f(z) = f_m(z) e^{if_a(z)}$, the regularity conditions imply:
\begin{eqnarray}
b(0)=1,f_{m}'(0)=f_{a}'(0)=0\label{bc2a}
\end{eqnarray}
The above equations are invariant under a  global phase of $f(z)$ so we can choose 
\begin{eqnarray}
f_{a}(0)=0\label{bca77}
\end{eqnarray}
For hyperbolic case the equations are determined by

\begin{eqnarray}
0 & = & b' -{ z(b^2 - z^2) \over b (f -\bar f)^2} f' \bar{f}' + {
 \omega (b^2 - z^2) \over b (f -\bar f)^2} (f \bar{f}'+ \bar{f} f')
+ {\omega^2 z |f|^2 \over b (f -\bar f)^2} \nonumber\\
0 & = & -f''
     - {z (b^2 + z^2) \over b^2 (f -\bar f)^2} f'^2 \bar{f}'
     + {2 \over (f -\bar f)} \left(\frac{1}{\bar f} 
       + { \omega (b^2 + z^2) \over 2b^2 (f -\bar f)} \right) \bar{f} f'^2,\nonumber \\&&
     + { \omega (b^2 + 2 z^2) \over b^2 (f -\bar f)^2} f f'
\bar{f}' 
  + {2 \over z} \left(-1+ { \omega z^2 (f +\bar f) \over (b^2 - z^2)
(f -\bar f)}\right.\nonumber \\&& 
+ \left.{\omega^2 z^4 |f|^2 \over b^2 (b^2 - z^2) 
(f -\bar f)^2}\right) f'- {\omega^2 z \over b^2 (f -\bar f)^2} f^2
\bar{f}' + \nonumber \\&&
{2 \omega \over (b^2 - z^2)} \left(-\frac{1}{2} - { \omega (f +\bar f )
\over 2(f -\bar f)}\right.
- \left.{\omega^2 z^2 |f|^2 \over 2b^2 (f -\bar f)^2}
\right) f.
\label{1fzeom321}
\end{eqnarray}

They are invariant under a constant scaling $f\rightarrow \lambda f$ 
and  applying regularity at the origin
we find that $f'(z=0)$ should vanish. Thus  the initial conditions for the hyperbolic case are:
\begin{eqnarray}
b(0)=1,f'(0)=0\label{bca}
\end{eqnarray}
Finally in the parabolic ansatz the equations of motion are invariant under arbitrary shifts of $f(z)$.

\section{Properties of the critical solutions} \label{sec:prop}

The properties of the solutions, the physical and geometrical behaviours of the solutions for elliptic case within details were explained in \cite{Hamade:1995jx,Eardley:1995ns}. Naturally, for the hyperbolic case the same properties are being held. In all equations we have  five singularities where $z=\pm 0$ corresponds to origin and we have dealt with them by making the regularity conditions. On the other hand, the point $z=\infty$ related to $t=0$. By change of variables and redefinition of the fields $f(z),b(z)$, one can show that \cite{hatefialvarez1307}
%we have shown in appendix A of 
 the equations remain regular there as well.

 \vskip.1in
%The singularities $b(z_{\pm})=\pm z_{\pm}$  are related to the backward
%(forward) light cones of the space-time. 

 The singularities $b(z_{\pm})=\pm z_{\pm}$  are the locations where the homothetic Killing vector is null, as explained in 
 \cite{AlvarezGaume:2011rk}. For $b(z_+)=z_+$ the solution must be
smooth within this surface and we need to have the  continuity of $f,b$ in this region. $b(z_{+}) =  z_{+}$ is related to the homothetic horizon and it is indeed a mere coordinate singularity \cite{Hirschmann:1995pr, AlvarezGaume:2011rk} so $\tau$ must be finite across it which gets interpreted as the finiteness of $f''(z)$ once $z\rightarrow z_+$. Another constraint comes from the fact that the vanishing of the divergent part of $f''(z)$ generates one complex valued constraint at $z_+$ that can be defined by  $G(b(z_+), f(z_+), f'(z_+)) = 0$ where the definitions of $G$ are given in Equations (49)-(51) of \cite{ours}. If we use regularity at $z=0$  as well as the residual symmetries, then we find out the initial conditions $b(0) = 1, f'(0) =0$ and the value of $f(0)$ is shown by

\begin{equation}
        f(0) = \left\{\begin{array}{l l l}
        x_0 & \text{elliptic}       & (0<x_0<1) \\
        i x_0 & \text{parabolic} & (0<x_0)\\
        1+i x_0 & \text{hyperbolic} & (0<x_0)
    \end{array}\right.
\end{equation}

where $x_0$ is a real parameter.  Thus, we have two constraints  as the real and imaginary parts of $G$ must be vanished  and two parameters $(\omega,x_0)$ to be known.

The entire solutions for hyperbolic case in four and five dimensions have been derived in \cite{Antonelli:2019dqv}. These solutions are obtained by making use of numerical integration from the equations of motion. For instance, for four dimensional elliptic case, just one solution is 
determined \cite{Eardley:1995ns,AlvarezGaume:2011rk} and it is given by
\begin{eqnarray}
\omega & = & 1.176 ,\nonumber\\
z_{ +} & = & 2.605 ,\nonumber\\
|f(0)| & = & 0.892,\nonumber\\
\label{last}
\end{eqnarray}

Using this new search methodology, we are able to explore the entire families of solutions for hyperbolic case in 4 dimension which are three cases called $\alpha,\beta,\gamma$ solutions. The $ \alpha$ solution is given by
\begin{eqnarray}
\omega & = & 1.362 ,\quad z_{ +} = 1.440, \quad 
Im f(0)= 0.708.\label{last}
\end{eqnarray}
The $\beta$ solution is determined by
\begin{eqnarray}
\omega & = & 1.003 ,\quad z_{ +} = 3.29,\quad 
Im f(0)= 0.0822.\label{last}
\end{eqnarray}
Finally $\gamma$ solution is explored to be
\begin{eqnarray}
\omega & = & 0.541 ,\quad z_{ +} = 8.44,\quad 
Im f(0)= 0.0059.\label{last}
\end{eqnarray}
%These solutions can be  \cite{Antonelli:2019dqv}.
% Now let us begin to  describe the methodology of exploring the estimate of the functions of $f(z), b(z)$ through various new techniques 
%such as Polynomial Regression Model, Kernel Regression Model and Local Regression Model.

\vskip.1in
 %  Making use of regularity at the origin and at $z_+$, applying the described initial conditions
%the critical solution will be determined.

%\begin{eqnarray}
% |v(0)|,\omega,z_+, |v(z_+)|
% \no%number
% \end{eqnarray}
\vskip.1in

%The methodology  for search for solutions that the authors in \cite{Eardley:1995ns} used, can be defined as follows. We first integrate out from the zero to positive values of $z_+$ and then we try to integrate in from $z_+$ into the origin.  Using Taylor expansion at $z_+$ and
%imposing regularity, matching the two solutions at an intermediate point (say $z=1$) and finally requesting continuity of functions and their first derivatives, one would be able to exactly explore the critical solution as was found for 4 dimension for elliptic case as follows:

%A recent research methodology for exploring critical solutions within details was explained in \cite{Antonelli:2019dqv}. For the sake of brevity we refer the interested reader to the section 4.1 of   \cite{Antonelli:2019dqv}.

%%%%%%%%%%%%%%%%%%%%%%%%%%%%%%%%%%%%%%%%%%%%%%%%%%%%%%%%%%%%%%%%%%%%%%%%%%%%%%%%%%%%%
\section{Statistical Estimation Methods}\label{sec:stat}
Throughout this section, we use the following notations to present the statistical estimation methods. 
Let $({\bf X},{\bf y})$ denote a multivariate random variable from a random sample of size $n$. 
Suppose ${\bf y}=(y_1\ldots,y_n)$ and ${\bf X}=({\bf x}_1,\ldots,{\bf x}_p)$ represent, respectively,
the vector of response  variable of size $n$ and $(n \times p)$ dimensional  design matrix 
with $p$ explanatory variables where $\text{rank}({\bf X})=p < n$.  

%%%%%%%%%%%%%%%%%%%%%%%%%%%%%%%%%%%%%%%%%%%%%%%%%%%%%%%%%%%%%%%%%%%%%%%%%%%%%%%%%%%%%
\subsection{Polynomial Regression Model}\label{sub:poly}
Linear regression models are among the most popular statistical methods for modelling data. 
One can address the relationship between response variable $y$ and the explanatory variables
 ${\bf x}_1,\ldots,{\bf x}_p$ by the linear regression model
 \begin{align} \label{reg}
 y_i = {\bf x}_i^\top {\bf \beta} + \epsilon_i, ~~ i=1,\ldots,n,
 \end{align}
 where  ${\bf \beta}$ is the unknown parameters (hence-after called coefficients)  of the model, 
 ${\bf x}_i^\top$ indicates the transpose of ${\bf x}_i$ and that 
 $\epsilon_i \overset{iid}{\sim} N(0,1); i=1,\ldots,n$; that is, the error terms  are independent and identically distributed 
 from standard normal distribution \cite{Harrell}.
 
 It is easily seen that linear regression model \eqref{reg}, as a parametric method, translates the prediction problem of response function $y=g(x)$ (as a function of explanatory variable $x$) to the estimation problem of the unknown parameters/coefficients of the model.
 Least square  (LS) method  \cite{Harrell} is one of the most common approaches to estimate the coefficients of model \eqref{reg}.
 Given the design matrix ${\bf X}$ and response vector ${\bf y}$ from $n$ observations, the least squares estimate of ${\bf \beta}$ 
   is given by 
   \begin{align} \label{ls}
   {\widehat{\bf\beta}}_{\text{LS}} = \underset{{\bf\beta}}{\arg \min} ~ ||{\bf y} - {\bf X}{\bf\beta}||_2^2,
   \end{align}
  where $||\cdot||_2^2$ denotes the $l_2$ norm. 
  It is easy to show that the solution to  \eqref{ls} is given by 
  \begin{align} \label{beta-hat_ls}
  {\widehat{\bf\beta}}_{\text{LS}} = ({\bf X}^\top {\bf X})^{-1} {\bf X}^\top {\bf y}.
  \end{align}
  Once model \eqref{reg} is trained, the response can be predicted at a new value ${\bf x}_{\text{new}}$ by
  \begin{align} \label{pred_ls}
  {\widehat y}_{\text{new}} = {\widehat g}({\bf x}_{\text{new}}) = {\bf x}_{\text{new}}^\top  {\widehat{\bf\beta}}_{\text{LS}}.  
  \end{align} 
  
  It is evident that the functional forms of $|f_0(z)|$, $\arg(f_0(z))$ and $b_0^2(z)-z^2$, as our underlying statistical population to be estimated are clearly nonlinear functions of 
 space-time. Hence, the simple linear regression model \eqref{reg} based on 
  $z$ is not flexible enough to estimate the nonlinearity of critical collapse functions. 
  One can employ polynomial regression model to deal with the nonlinear critical collapse  functions. 
  Polynomial regression model enables us
   to incorporate the higher orders of explanatory variable $x$ to approximate better the nonlinear response function $y=g(x)$.
   Polynomial regression model of order $l$ is given by
   \begin{align} \label{poly}
   y_i = \sum_{j=0}^{l} x_i^j \beta_j + \epsilon_i, 
   \end{align} 
   where ${\bf\beta}=(\beta_0,\ldots,\beta_l)$ represent the unknown coefficients of the model. 
   Note that, we only focus on the main effects of explanatory variables (and their higher orders) in estimating the critical functions. First, in the estimating of the critical functions, there is only a single explanatory variable -- that is the space-time $z$. Hence, no interaction term is defined in the regression models. When there is a single explanatory variable in the regression, the higher orders of the explanatory variable can very well accommodate the nonlinearity of the  population. Finally, the interaction terms typically contribute to the refinement of the estimates at the price of introducing more parameters in the model and reducing the degree of freedom in the estimation. For the above reasons, throughout these manuscript, we do not include the interaction terms in the statistical models.
   
   Polynomial regression model, as a special case of model \eqref{reg}, can be written as a linear model again by
 \begin{align*} 
  {\bf y} = {\bf Z} {\bf \beta} + \epsilon, 
 \end{align*}
 where the columns of matrix ${\bf Z}$ are the copies of explanatory variable $x_i$ taken to various powers $j=0,\ldots,l$. 
 Similarly, from least squares method \eqref{ls}, the polynomial regression at ${\bf x}_{\text{new}}$ is produced by 
 \begin{align} \label{pred_poly}
  {\widehat y}_{\text{new}} = {\widehat g}({\bf x}_{\text{new}}) = {\bf x}_{\text{new}}^\top ({\bf Z}^\top {\bf Z})^{-1} {\bf Z}^\top {\bf y}.  
  \end{align}
  Polynomial regression model provides a flexible  solution to estimate a nonlinear function at the price of higher orders of explanatory 
  variable $x$ in the model. Therefore, the estimation performance of the polynomial proposal depends on the order of the polynomial regression $l$.
  In Section \ref{sec:num}, we perform a cross validation to select the best order of the polynomial estimators. 
  
    %%%%%%%%%%%%%%%%%%%%%%%%%%%%%%%%%%%%%%%%%%%%%%%%%%%%%%%%%%%%%%%%%%%%%%%%%%%%%%%%%%%%%
\subsection{Kernel Regression Model} \label{sub:kernel}
 Linear regression and polynomial regression models translate the estimation problem of the response 
 function $y=g(x)$ to the estimation problem of parameters  ${\bf\beta}$. Non-parametric regression 
 models can be considered as another approach to estimate the nonlinear critical collapse functions.  
 Kernel regression model is one of the most common non-parametric estimation methods.  
 Kernel regression approximates the response function at new observation ${x}_{\text{new}}$ by a 
 weighted average of observed responses in a neighbourhood of ${x}_{\text{new}}$.
 A kernel function is non-negative symmetric function around the origin (i.e, the centre of the neighbourhood).
 Kernel function is typically re-scaled to result in a legitimate probability density function in each neighbourhood. 
 There are various  choices for Kernel function; however, in this manuscript, we focus only on Epanechnikov kernel \cite{Cleveland} given by
 \begin{align}\label{epan}
 K_h(x) = 3/4\left(1-(x/h)^2\right) I(|x/h| \le 1),
 \end{align}
 where $h$ denotes the bandwidth parameter of the kernel function $K$ and $I$ is an indicator function such 
 that $I(u)=1$ if $u$=true otherwise $I(u)=0$. It is at once apparent that the kernel function \eqref{epan} 
 tunes the width of neighbourhoods based on the bandwidth parameter $h$. 
 When an observation falls out of  the bandwidth, the kernel function assigns very small weight to the 
 observation to reduce its impact on the function estimate. 
 
 Given a training data set of size $n$ with explanatory variable $x$ and response $y$, the Nadaraya--Watson 
 kernel regression estimate \cite{Nadaraya,Watson} the response function at new observation $x_{\text{new}}$ as
 \begin{align}\label{pred_kernel}
   {\widehat y}_{\text{new}} = {\widehat g}({\bf x}_{\text{new}}) = \sum_{i=1}^{n} \frac{K_h\left(|x_i -x_{\text{new}}| \right) y_i}{\sum_{r=1}^{n} 
   K_h\left(|x_{r} -x_{\text{new}}| \right)},
 \end{align}
  where $K_h(\cdot)$ is obtained from \eqref{epan}.
  
  Note that the kernel regression estimator \eqref{pred_kernel} deals with the nonlinearity of 
  response function at the price of  selecting the bandwidth parameter.  To this end, when small $h$  
  is selected, the weights assigned by kernel regression estimator are more concentrated around the 
  new observation (i.e., shorter neighbourhoods will be  declared). In contrast, when large $h$ is selected,
   the weights will be more spread out and consequently wider neighbourhoods are declared. Hence, for 
   a given training data, we carry out a cross validation to find out the optimal bandwidth.  
 
   %%%%%%%%%%%%%%%%%%%%%%%%%%%%%%%%%%%%%%%%%%%%%%%%%%%%%%%%%%%%%%%%%%%%%%%%%%%%%%%%%%%%%
\subsection{Local Regression Model} \label{sub:loc}
As another approach, we study the local regression model to earn the curvature of the response function $y=g(x)$.
Local regression, as a semi-parametric approach, combines the parametric advantages of the polynomial regression and non-parametric properties of the kernel regression in estimating the response function. 
Local regression can be intuitively explained by the Taylor  expansion of $g(x)$ around ${x}_{\text{new}}$ as follows:
\[
g({x}_{\text{new}}+\delta) = g({x}_{\text{new}}) + \delta \frac{\partial g}{\partial x} ({x}_{\text{new}}) + 
\frac{\delta^2}{2}  \frac{\partial^2 g}{\partial x^2} ({x}_{\text{new}}) + O(\delta^3),
\]  
where kernel regression  can be viewed as an estimator which only utilizes the constant term to approximate $g(x)$.
Local regression method exploits  the high orders of $x$ by polynomial regression and then estimates  the coefficients (of the polynomial regression) by the kernel regression in each neighbourhood. 

Given a training data set of size n with explanatory variable $x$ and response variable $y$, the local regression estimator \cite{Cleveland} of order $l$ at  
${\bf x}_{\text{new}}$ is given by
\begin{align} \label{pred_local}  
{\widehat y}_{\text{new}} = {\widehat g}({\bf x}_{\text{new}}) = \sum_{j=0}^{l} x_i^j \widehat{\beta_j}({x}_{\text{new}}),
\end{align}
where the vector of coefficients estimate ${\widehat{\bf\beta}}({x}_{\text{new}})= \left(\widehat{\beta_0}({x}_{\text{new}}),\ldots, \widehat{\beta_l}({x}_{\text{new}}) \right)$ is obtained as a solution to:
\begin{align} \label{beta_local} 
{\widehat{\bf\beta}}({x}_{\text{new}}) = \underset{{\bf\beta}}{\arg \min} 
\left\{ \sum_{i=1}^{n} K_h\left(|x_i -x_{\text{new}}| \right) \left(y_i - \sum_{j=0}^{l} x_i^j \beta_j\right)^2
\right\},
\end{align}
where $K_h(\cdot)$ obtained from \eqref{epan}. 

 \begin{figure}
\includegraphics[width=1.2\textwidth,center]{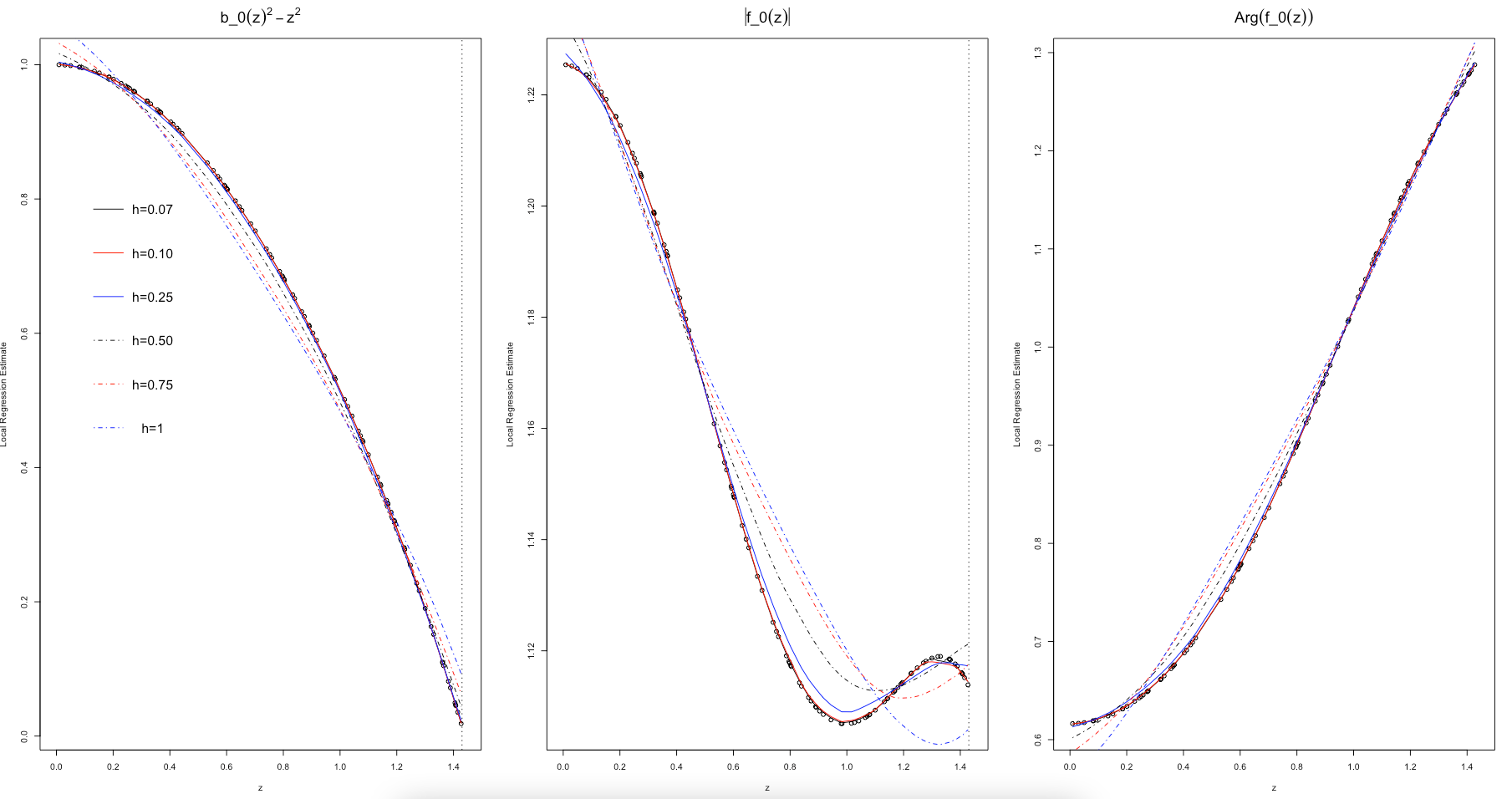}
\caption{\footnotesize{The estimates of critical collapse functions corresponding to $\alpha$ solution of hyperbolic case based on local 
regression method of order $l=1$ with bandwidth parameters {$h=\{0.07,0.10,0.25,0.50,0.75,1\}$} 
based on a training sample of size $n=100$.}}
 \label{hy_al_local}
\end{figure}

The estimator \eqref{pred_local} can be used to estimate the 
response function at any value of ${x}_{\text{new}}$. Accordingly,  the estimator \eqref{pred_local} enables us to approximate 
 the response function $y=g(x)$ throughout the domain. Note that the local regression smoother \eqref{pred_local} requires 
 two tuning parameters. These tuning parameters include  the bandwidth parameter  of kernel part $h$ and the order of the polynomial part $l$.

  \begin{figure}
\includegraphics[width=1.2\textwidth,center]{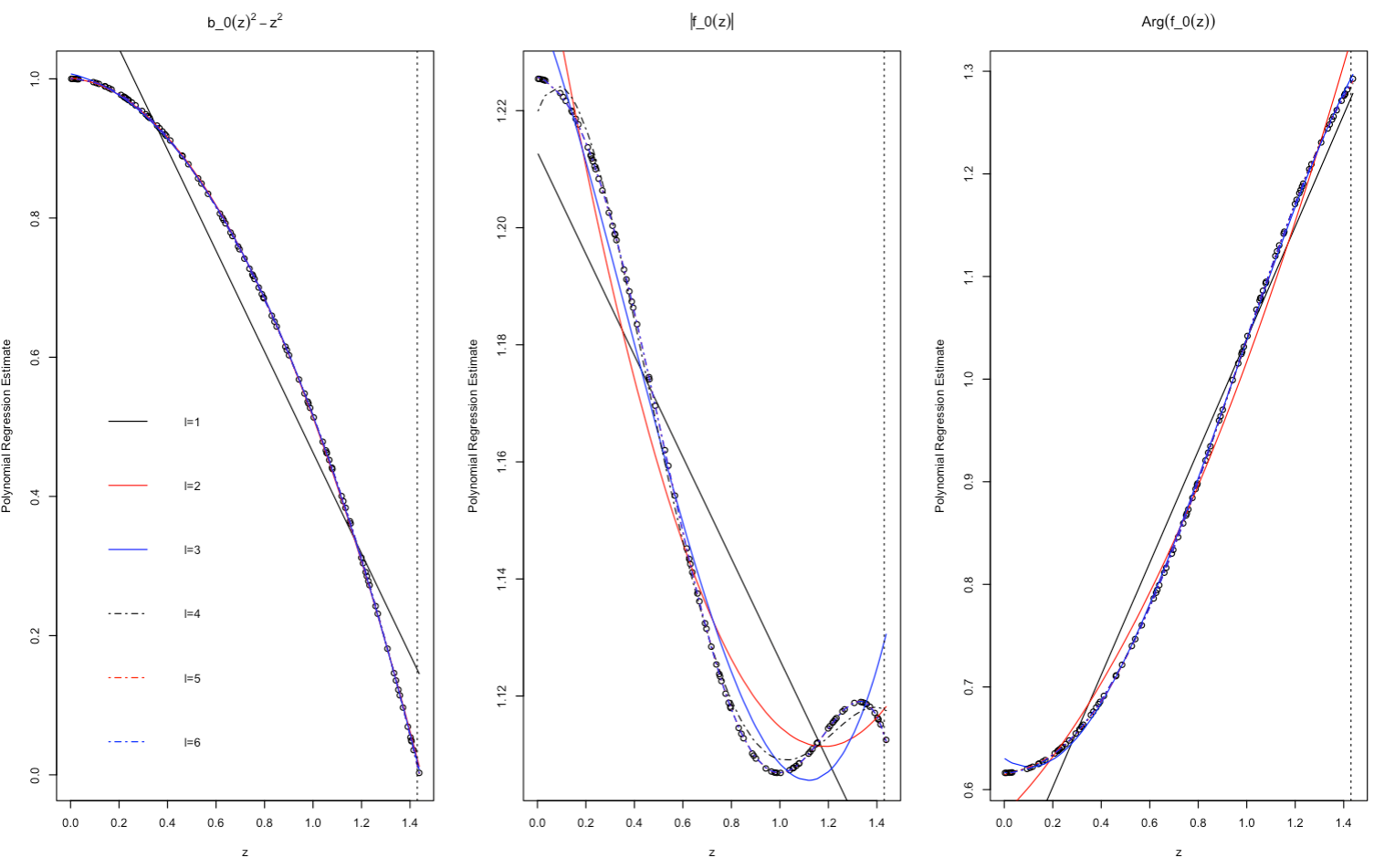}
\caption{\footnotesize{The estimates of critical collapse functions corresponding to $\alpha$ solution of hyperbolic case based on polynomial 
regression method of orders $l=\{1,\ldots,6\}$ based on a training sample of size $n=100$.}}
 \label{hy_al_poly}
\end{figure}

%%%%%%%%%%%%%%%%%%%%%%%%%%%%%%%%%%%%%%%%%%%%%%%%%%%%%%%%%%%%%%%%%%%%%%%%%%%%%%%%%%%%%
\section{Numerical Studies} \label{sec:num}
  
 R. Antonelli and E. Hatefi  in \cite{ours} recently studied the black hole solutions of axion-dilaton system in elliptic and hyperbolic cases in four and five dimension.
 Through the numerical  optimization of \cite{Eardley:1995ns}, they found only one solution to equations of motion for 4 dimension of elliptic case.
 %Accordingly, they show that there is there exits a unique elliptic solution. They also show that there must 
 %be at least three critical solutions   in hyperbolic scenario. 
 As discussed in \cite{Antonelli:2019dqv}, the unperturbed critical collapse functions play a key role in the 
 location of the critical solutions and critical exponents. Despite the importance of these unperturbed critical collapse functions, a little information is known in the literature about the structure and closed form of these functions.  
 It is, thus, of high importance for researchers to estimate numerically the functional form of these 
 unperturbed functions so that the critical solutions and critical exponents as well as mass of Black Holes and universality of Choptuik exponents will be more tractable.  In this section, we employ non-linear statistical methods including polynomial regression, non-parametric kernel regression and local polynomial regression methods to estimate
 the functional forms of the unperturbed critical collapse functions. 
  \begin{table}[ht]
\centering
\footnotesize{\begin{tabular}{cccc}
\hline
   Elliptic &                  &Hyperbolic        &    \\ 
     \cline{2-4} 
            &$\alpha$-solution &$\beta$-solution  & $\gamma$-solution \\ 
\hline
 0.00370  (.10) & 0.00040  (.060) & 0.00681  (.140) & 0.13307  (.29) \\ 
 0.00438  (.11) & 0.00043  (.062) & 0.00702  (.142) & 0.13375  (.30) \\ 
 0.00521  (.12) & 0.00045  (.064) & 0.00723  (.144) & 0.13450  (.31) \\ 
 0.00608  (.13) & 0.00048  (.066) & 0.00743  (.146) & 0.13532  (.32) \\ 
 0.00693  (.14) & 0.00050  (.068) & 0.00763  (.148) & 0.13909  (.33) \\ 
 0.00780  (.15) & 0.00053  (.070) & 0.00782  (.150) & 0.14370  (.34) \\ 
 0.00870  (.16) & 0.00055  (.072) & 0.00801  (.152) & 0.14677  (.35) \\ 
 0.00980  (.17) & 0.00058  (.074) & 0.00820  (.154) & 0.14910  (.36) \\ 
 0.01093  (.18) & 0.00060  (.076) & 0.00839  (.156) & 0.15134  (.37) \\ 
 0.01203  (.19) & 0.00063  (.078) & 0.00861  (.158) & 0.15342  (.38) \\ 
 \hline
\end{tabular}
\caption{\footnotesize{ The $\sqrt{\text{MSE}}$ of local regression method with $l=1$ and bandwidth $h$ (presented in parenthesis) in estimating 
the critical collapse response function $g(z)=b_0^2(z)-z^2$ in elliptic and hyperbolic cases based on a training sample of size $n=100$.}}
\label{b_local}}
\end{table}

\begin{table}[ht]
\centering
\footnotesize{\begin{tabular}{cccc}
\hline
   Elliptic &                  &Hyperbolic        &    \\ 
     \cline{2-4} 
            &$\alpha$-solution &$\beta$-solution  & $\gamma$-solution \\ 
\hline
 0.00052  (.10) & 0.00021  (.060) & 0.00024  (.140) & 0.00005  (.29) \\ 
 0.00059  (.11) & 0.00022  (.062) & 0.00025  (.142) & 0.00006  (.30) \\ 
 0.00069  (.12) & 0.00024  (.064) & 0.00025  (.144) & 0.00006  (.31) \\ 
 0.00080  (.13) & 0.00025  (.066) & 0.00026  (.146) & 0.00006  (.32) \\ 
 0.00090  (.14) & 0.00026  (.068) & 0.00027  (.148) & 0.00006  (.33) \\ 
 0.00099  (.15) & 0.00027  (.070) & 0.00027  (.150) & 0.00006  (.34) \\ 
 0.00109  (.16) & 0.00028  (.072) & 0.00028  (.152) & 0.00006  (.35) \\ 
 0.00121  (.17) & 0.00029  (.074) & 0.00028  (.154) & 0.00006  (.36) \\ 
 0.00132  (.18) & 0.00030  (.076) & 0.00029  (.156) & 0.00006  (.37) \\ 
 0.00143  (.19) & 0.00031  (.078) & 0.00029  (.158) & 0.00006  (.38) \\ 
 \hline
\end{tabular}
\caption{\footnotesize{The  $\sqrt{\text{MSE}}$ of local regression method with $l=1$ and bandwidth $h$ (presented in parenthesis) in estimating 
the critical collapse response function $g(z)=|f_0(z)|$ in elliptic and hyperbolic cases based on a training sample of size $n=100$.}}
\label{absf_local}}
\end{table}
 Using the optimization techniques of \cite{Eardley:1995ns} a numerical search is carried out to find 
 the critical solution on various intervals in the domain of forward singularity ( $[0,z_{+}]$). Accordingly, they showed that there 
 was a unique critical solution in elliptic space. This results in the interval $[0,2.5]$ as the 
 domain of the critical collapse functions in elliptic space in four dimension, where this unique solution was also confirmed in \cite{ours}. 
 Similarly,  R. Antonelli and E. Hatefi \cite{ours} explored three solutions (say $\alpha, \beta$ and $\gamma$ 
  critical solutions) to the equation of motion in hyperbolic case. This leads to three corresponding domains, including 
 $[0,1.44], [0,3.30]$ and $[0,8.45]$ for the unperturbed functions. In a similar vein to  \cite{ours}, we carried out the
 optimization search and obtained 2000 observations  from the critical functions $b_0^2-z^2$, $|f_0|$ and $\arg f_0$ 
 in all elliptic and hyperbolic domains. These observations were treated as the (unknown) underlying statistical populations to be estimated. 
 
 \begin{table}[ht]
\centering
\footnotesize{\begin{tabular}{cccc}
\hline
   Elliptic &                  &Hyperbolic        &    \\ 
     \cline{2-4} 
            &$\alpha$-solution &$\beta$-solution  & $\gamma$-solution \\ 
\hline
 0.00020  (.10) & 0.00026  (.060) & 0.00074  (.140) & 0.00138  (.29) \\ 
 0.00024  (.11) & 0.00028  (.062) & 0.00077  (.142) & 0.00138  (.30) \\ 
 0.00028  (.12) & 0.00030  (.064) & 0.00079  (.144) & 0.00138  (.31) \\ 
 0.00032  (.13) & 0.00032  (.066) & 0.00082  (.146) & 0.00138  (.32) \\ 
 0.00037  (.14) & 0.00034  (.068) & 0.00084  (.148) & 0.00140  (.33) \\ 
 0.00041  (.15) & 0.00036  (.070) & 0.00086  (.150) & 0.00143  (.34) \\ 
 0.00047  (.16) & 0.00037  (.072) & 0.00088  (.152) & 0.00145  (.35) \\ 
 0.00053  (.17) & 0.00039  (.074) & 0.00090  (.154) & 0.00146  (.36) \\ 
 0.00060  (.18) & 0.00041  (.076) & 0.00092  (.156) & 0.00147  (.37) \\ 
 0.00067  (.19) & 0.00043  (.078) & 0.00094  (.158) & 0.00148  (.38) \\ 
 \hline
\end{tabular}
\caption{\footnotesize{The  $\sqrt{\text{MSE}}$ of local regression method with $l=1$ and bandwidth $h$ (presented in parenthesis) in estimating 
the critical collapse response function $g(z)=\arg f_0(z)$ in elliptic and hyperbolic cases based on a training sample of size $n=100$.}}
\label{argf_local}}
\end{table}

 For each observation in the population, we generated four characteristics from the valid domain of unperturbed critical solution of $ [0,z_{+}]$.    
 These characteristics  include the realizations of critical functions $b_0^2-z^2$, $|f_0|, \arg f_0$ and $z$. 
 In the statistical analysis, we considered space-time $z$ as the single explanatory variable $(x)$ and the realization of the critical collapse functions $b_0^2-z^2$, $|f_0|, \arg f_0$ as the responses (observed from the corresponding critical function) in the regression models. We fitted one regression model for each critical function.
%We treated $z$ as the explanatory  variable and observations of $b_0^2-z^2$, $|f_0|, \arg f_0$  as the responses received from these critical functions 
% in the statistical analysis. 
 We generated independently (i.e., with replacement) training samples of size $n=100$ from each population. 
 For the validation of the estimation, we generated (independent from the training data) test data $(x_{test},y_{test})$ of size $n=100$ from the entire domains of the critical functions. 
As described in Section \ref{sec:stat}, to estimate the critical function by the polynomial regression,  we first applied equation \eqref{beta-hat_ls} to the training data and estimated the coefficients of the model. Using the estimated coefficients ${\widehat\beta}$ and \eqref{pred_ls}, we then predicted the response of the critical function $(\widehat{y}_{test})$ at $x_{test}$.
For Kernel regression method, we applied equation \eqref{pred_kernel} to the training data and predicted the value of  the critical function $(\widehat{y}_{test})$ at test point $x_{test}$.
According to the definition of the local regression, the coefficients of the model are treated as the functions of the test data. Hence, we used the training data and estimated the coefficients  using \eqref{beta_local}. 
From \eqref{pred_local} and the estimated the coefficients  ${\widehat\beta(x_{test})}$, we then predicted the critical collapse function $(\widehat{y}_{test})$ at $x_{test}$. 
We finally implemented the above prediction procedures sequentially for all the points in test data to estimate all the critical collapse functions over their entire domains.

% As described in Section \ref{sec:stat}, the training samples were  used to train the polynomial regression, non-parametric kernel regression  and local regression models for each critical collapse function. Independent from the training samples, we generated a test data $({\bf x}_{test},{\bf y}_{test})$ of size $n=100$.
% Using the trained regression estimators, we predicted the response functions at each level of ${\bf x}_{test}$, say ${\widehat{\bf y}}_{test}$.
 
 To assess the accuracy of the proposed estimators, we use the measure of square root of the mean squared errors $\sqrt{\text{MSE}}$ as follows
 \[
 \sqrt{\text{MSE}} = \left( 1/n \sum_{i=1}^{n} ({\widehat{y}}_{test,i} - {{y}}_{test,i})\right)^{1/2},
 \]
Note that  the trained model will be more  accurate in estimating the critical collapse response function when $\sqrt{\text{MSE}}$ is small. 
 To investigate the impact of tuning parameters on the performance of the estimators, similar to above, we generated training data and validation data of sizes $n=100$ from the population and computed the $\sqrt{\text{MSE}}$s of estimates of critical collapse functions for $h=\{0.01,0.02,\ldots,0.5\}$ and $l=\{1,2,\ldots,10\}$. 
 
\begin{figure}
\includegraphics[width=1.2\textwidth,center]{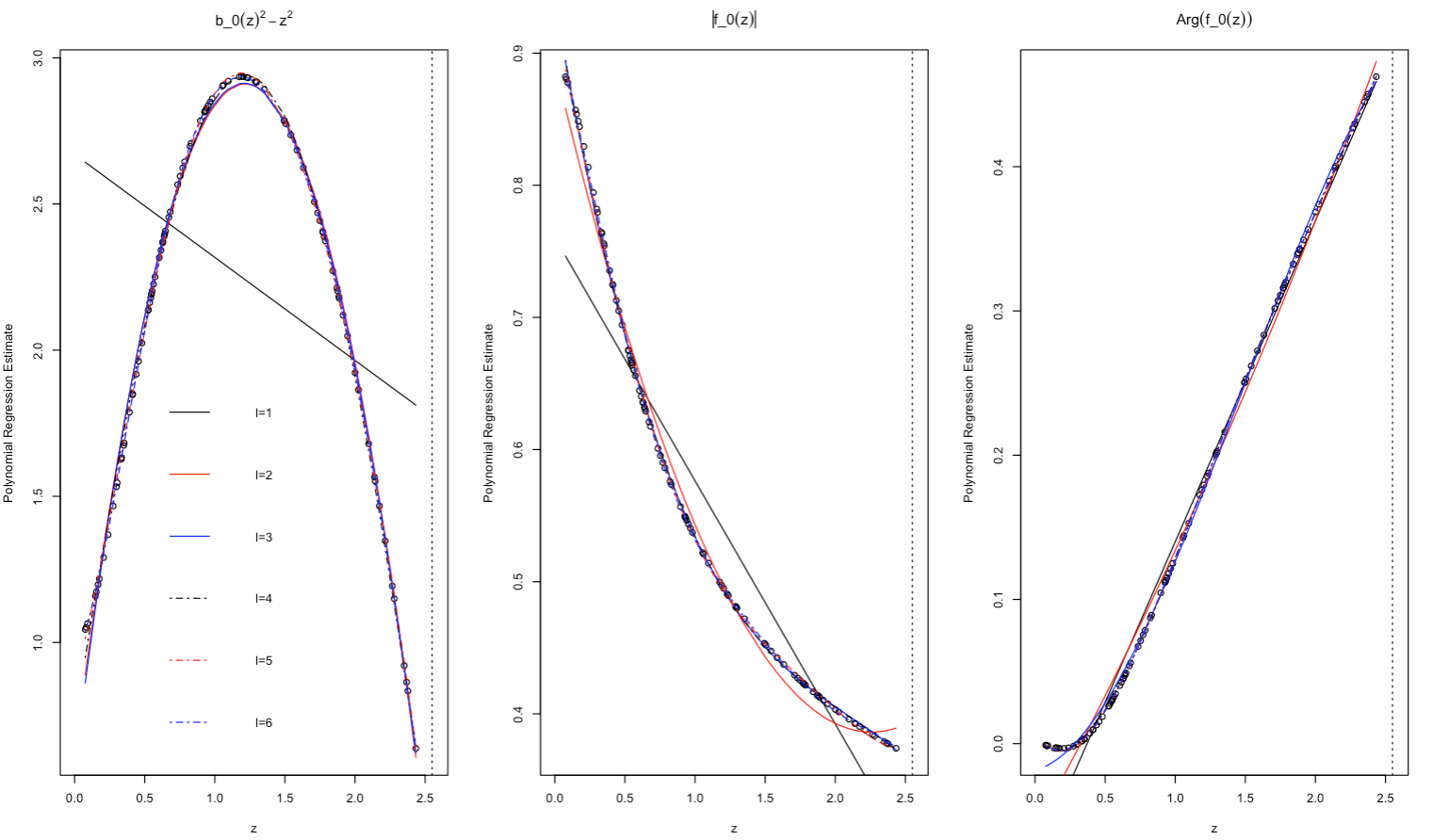}
\caption{\footnotesize{The estimates of critical collapse functions based on polynomial 
regression method of orders $l=\{1,\ldots,6\}$ in elliptic case based on a training sample of size $n=100$.}}
 \label{elip_poly}
\end{figure}

\begin{figure}
\includegraphics[width=1.2\textwidth,center]{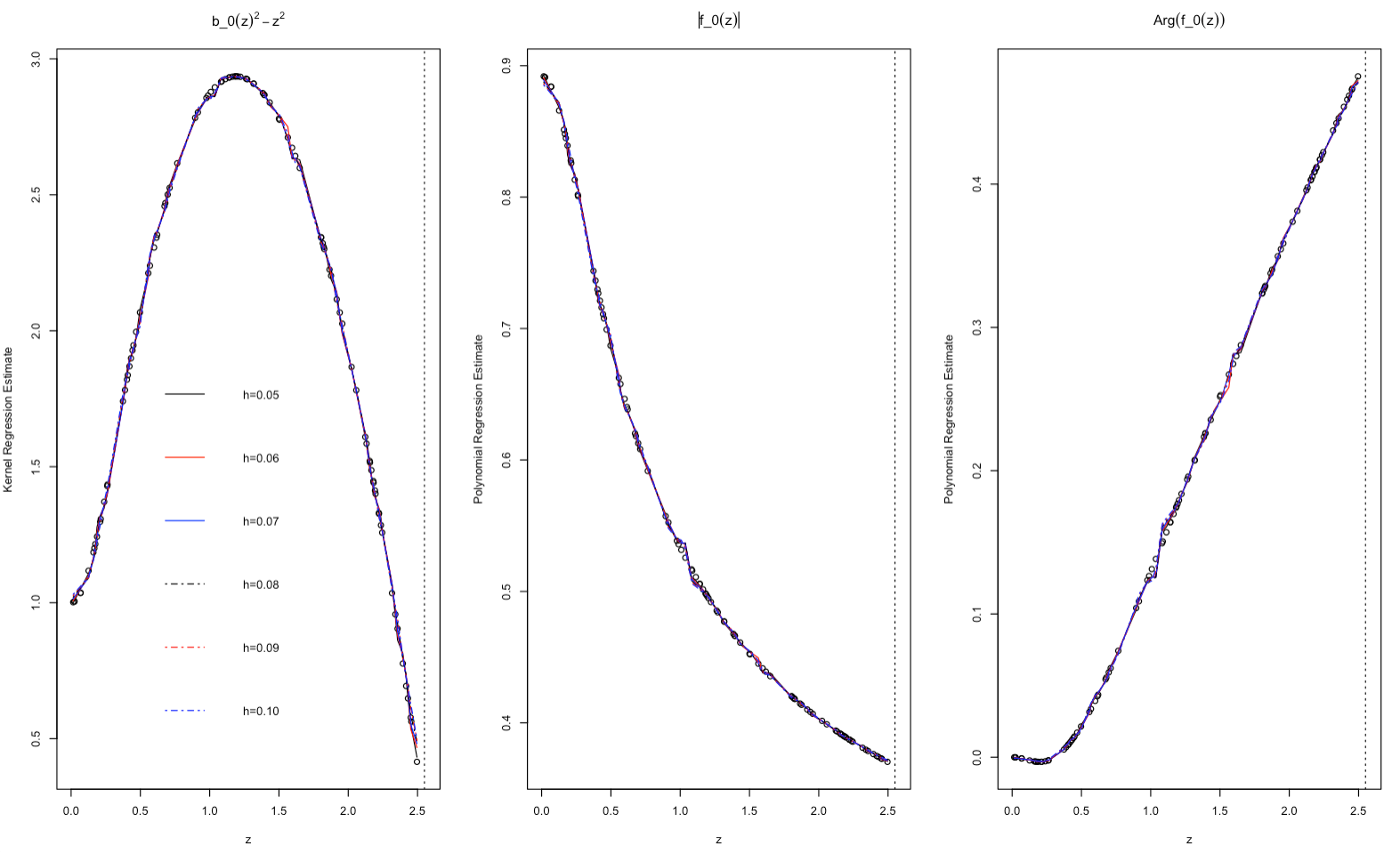}
\caption{\footnotesize{The estimates of critical collapse functions based on kernel 
regression method with bandwidth parameters $h=\{0.05,0.06,0.07,0.08,0.09,0.10\}$ in elliptic case 
based on a training sample of size $n=100$.}}
 \label{elip_kernel}
\end{figure}

\begin{figure}
\includegraphics[width=1.2\textwidth,center]{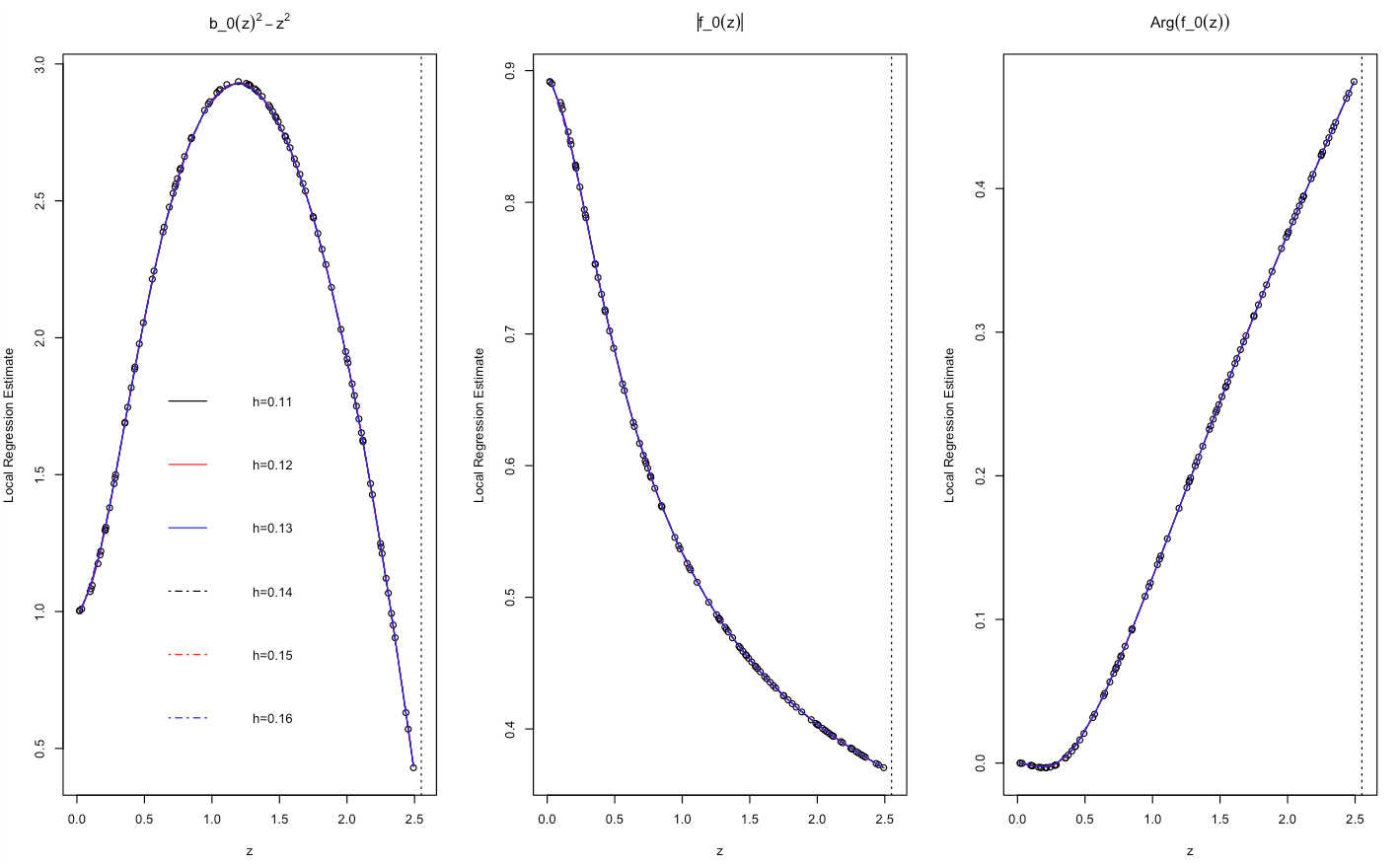}
\caption{\footnotesize{The estimates of critical collapse functions based on local 
regression method of order $l=1$ with bandwidth parameters $h=\{0.11,0.12,0.13,0.14,0.15,0.16\}$ in elliptic case 
based on a training sample of size $n=100$.}}
 \label{elip_local}
\end{figure}

\begin{figure}
\includegraphics[width=1.2\textwidth,center]{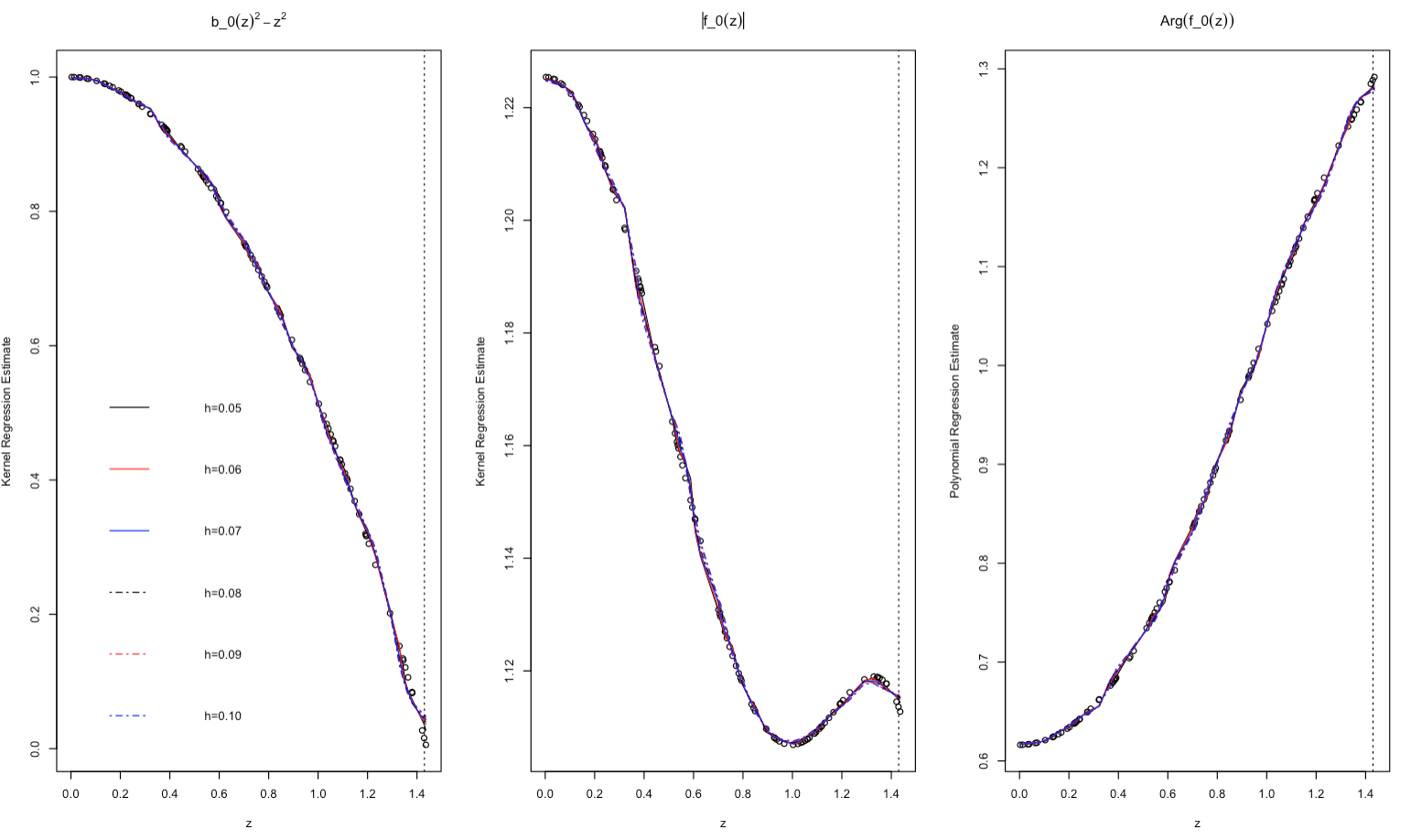}
\caption{\footnotesize{The estimates of critical collapse functions corresponding to $\alpha$ solution of hyperbolic case based on kernel 
regression method with bandwidth parameters $h=\{0.05,0.06,0.07,0.08,0.09,0.10\}$ 
based on a training sample of size $n=100$.}}
 \label{hy_al_kernel}
\end{figure}

\begin{figure}
\includegraphics[width=1.2\textwidth,center]{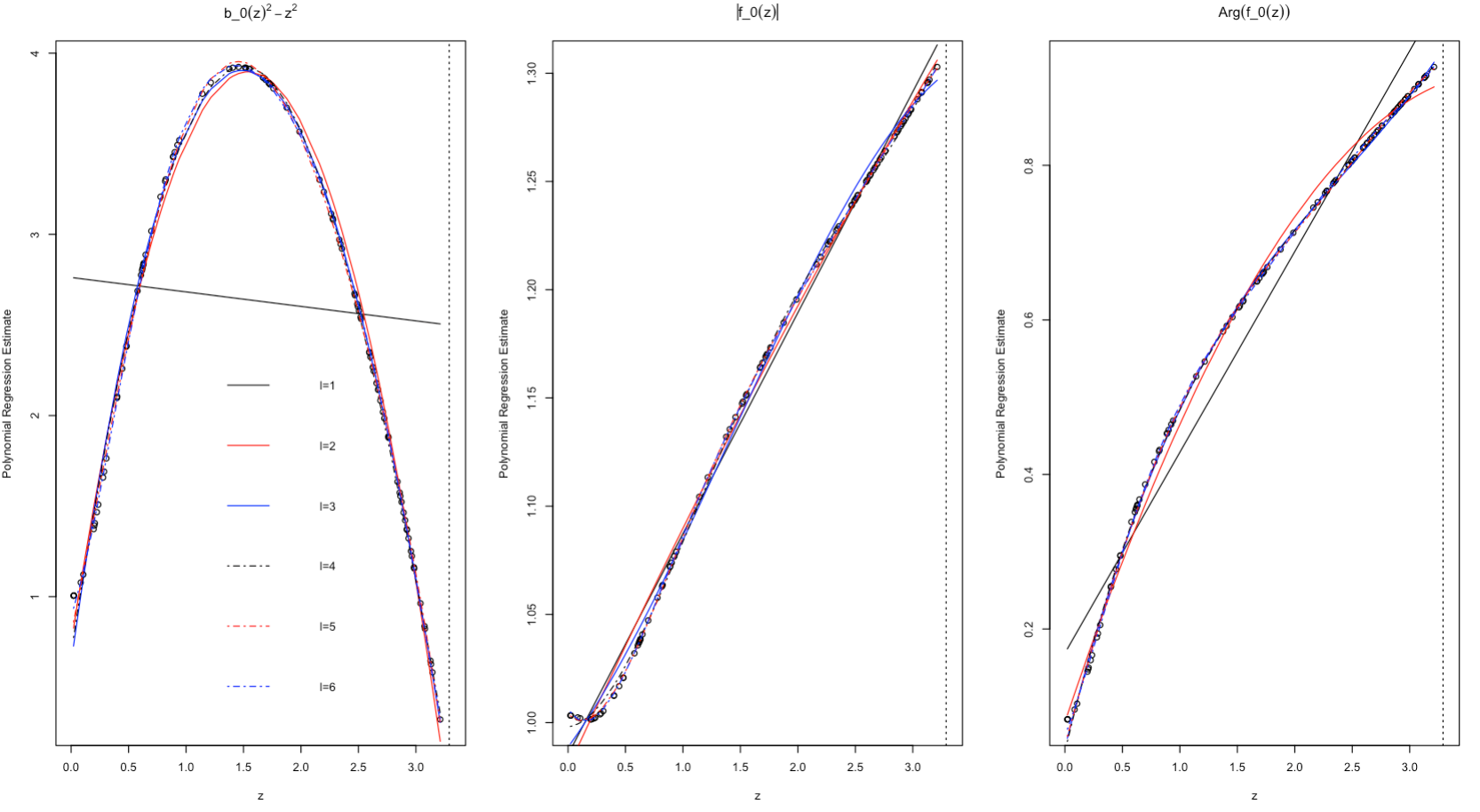}
\caption{\footnotesize{The estimates of critical collapse functions corresponding to $\beta$ solution of hyperbolic case based on polynomial 
regression method of orders $l=\{1,\ldots,6\}$ based on a training sample of size $n=100$.}}
 \label{hy_be_poly}
\end{figure}

\begin{figure}
\includegraphics[width=1.2\textwidth,center]{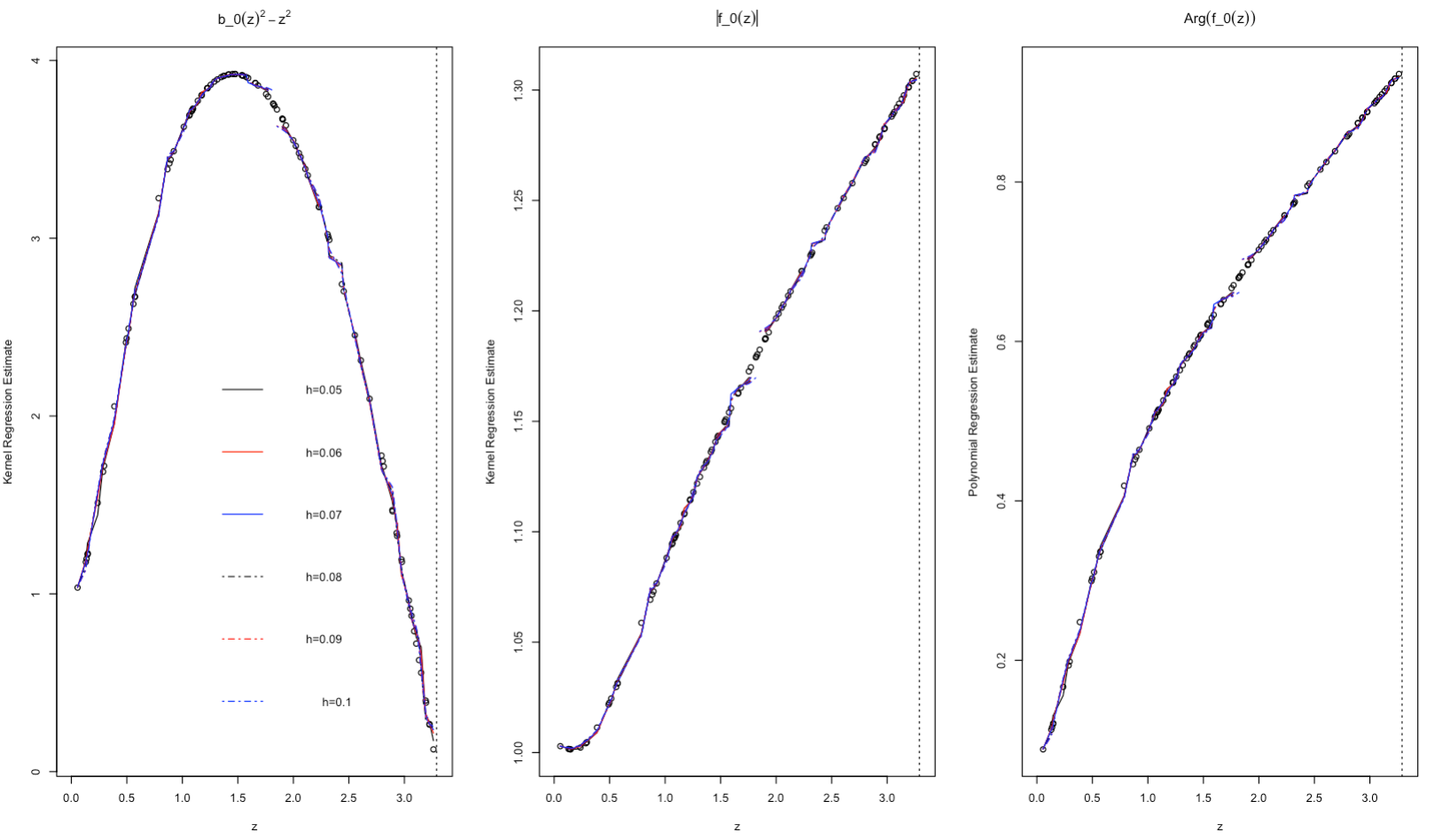}
\caption{\footnotesize{The estimates of critical collapse functions corresponding to $\beta$ solution of hyperbolic case based on kernel 
regression method with bandwidth parameters $h=\{0.05,0.06,0.07,0.08,0.09,0.10\}$ 
based on a training sample of size $n=100$.}}
 \label{hy_be_kernel}
\end{figure}

\begin{figure}
\includegraphics[width=1.2\textwidth,center]{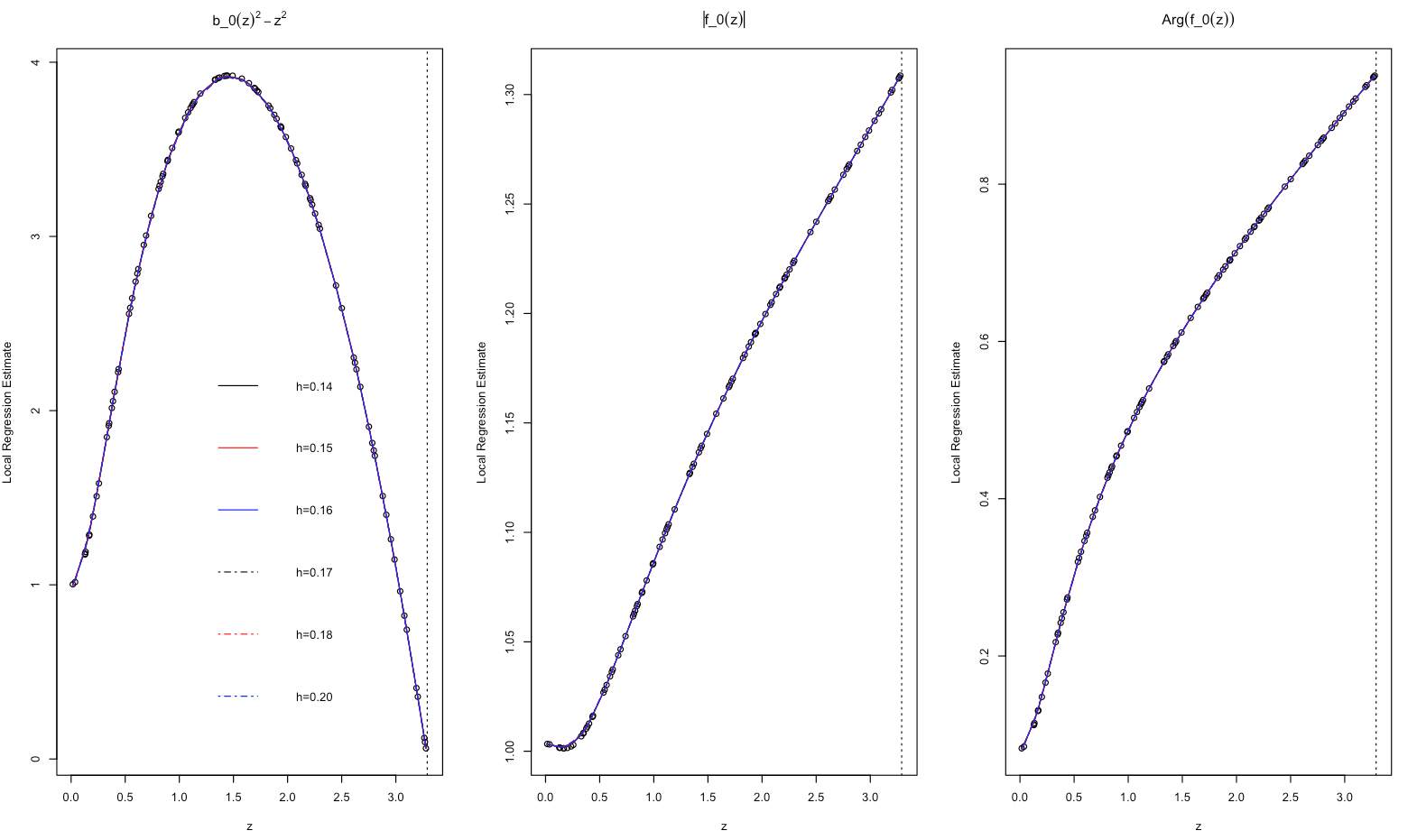}
\caption{\footnotesize{The estimates of critical collapse functions corresponding to $\beta$ solution of hyperbolic case based on local 
regression method of order $l=1$ with bandwidth parameters $h=\{0.14,0.15,0.16,0.17,0.18,0.20\}$ 
based on a training sample of size $n=100$.}}
 \label{hy_be_local}
\end{figure}

\begin{figure}
\includegraphics[width=1.2\textwidth,center]{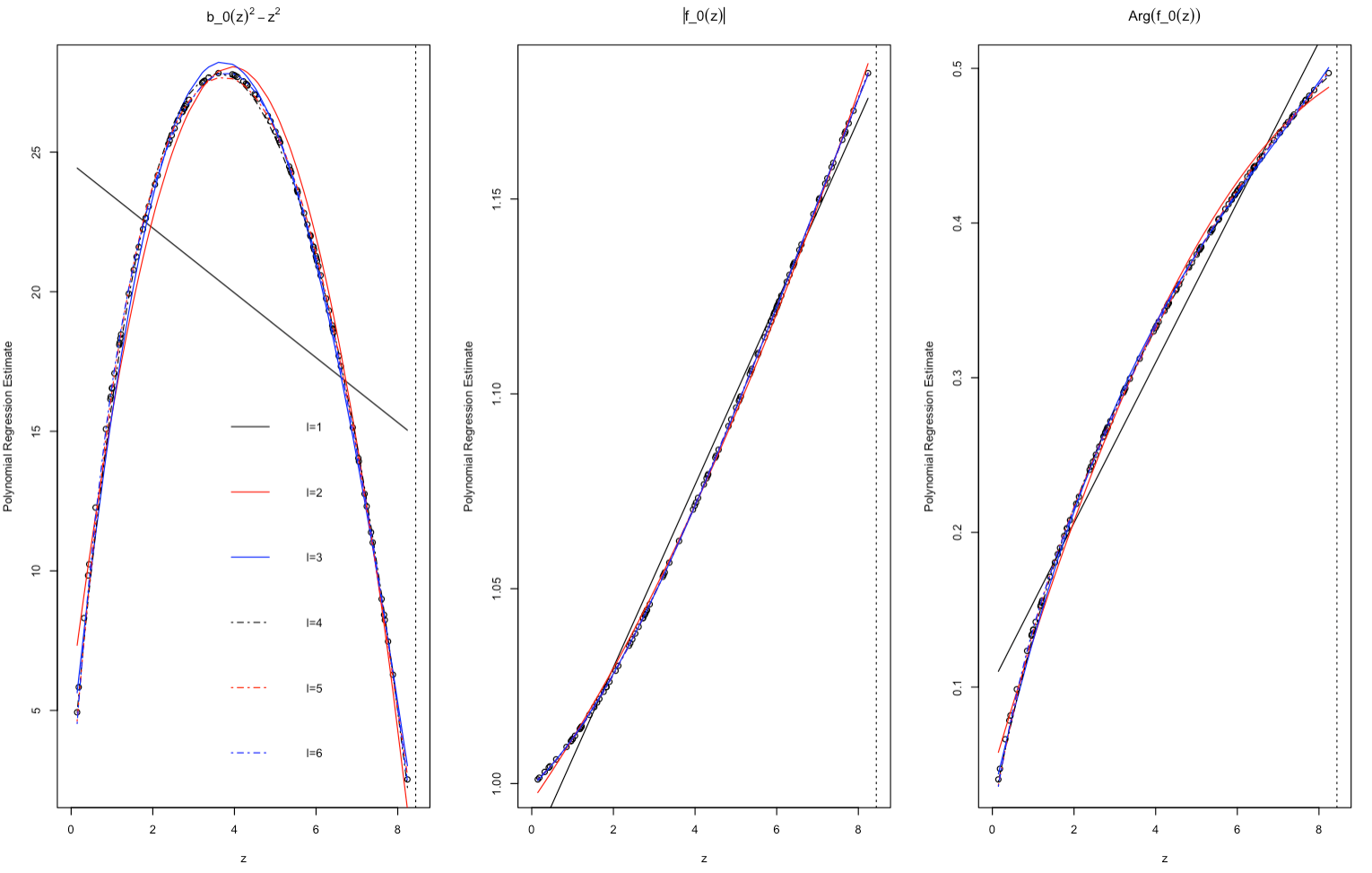}
\caption{\footnotesize{The estimates of critical collapse functions corresponding to $\gamma$ solution of hyperbolic case based on polynomial 
regression method of orders $l=\{1,\ldots,6\}$ based on a training sample of size $n=100$.}}
 \label{hy_ga_poly}
\end{figure}

\begin{figure}
\includegraphics[width=1.2\textwidth,center]{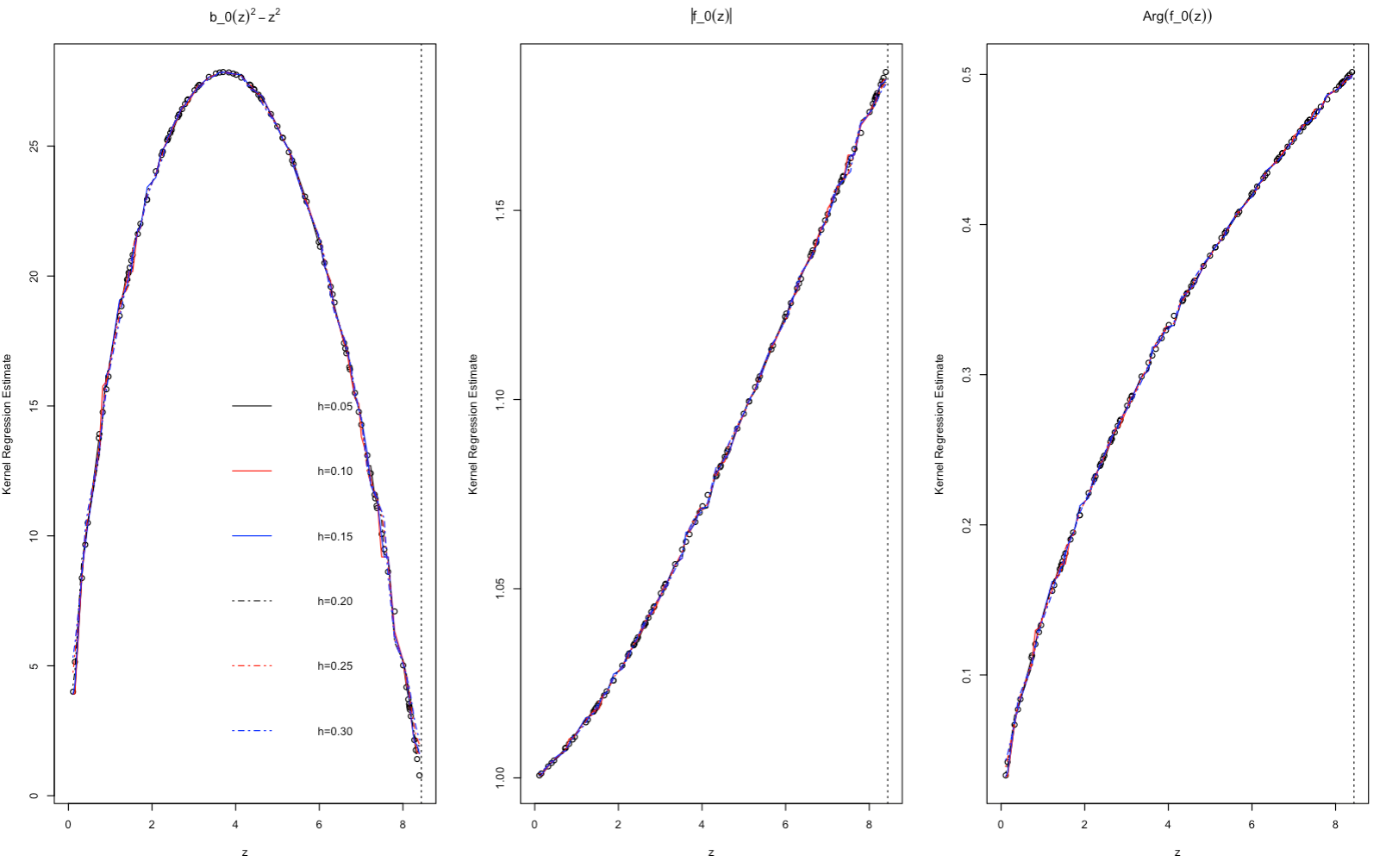}
\caption{\footnotesize{The estimates of critical collapse functions corresponding to $\gamma$ solution of hyperbolic case based on kernel 
regression method with bandwidth parameters $h=\{0.05,0.10,0.15,0.20,0.25,0.30\}$ 
based on a training sample of size $n=100$.}}
 \label{hy_ga_kernel}
\end{figure}

\begin{figure}
\includegraphics[width=1.2\textwidth,center]{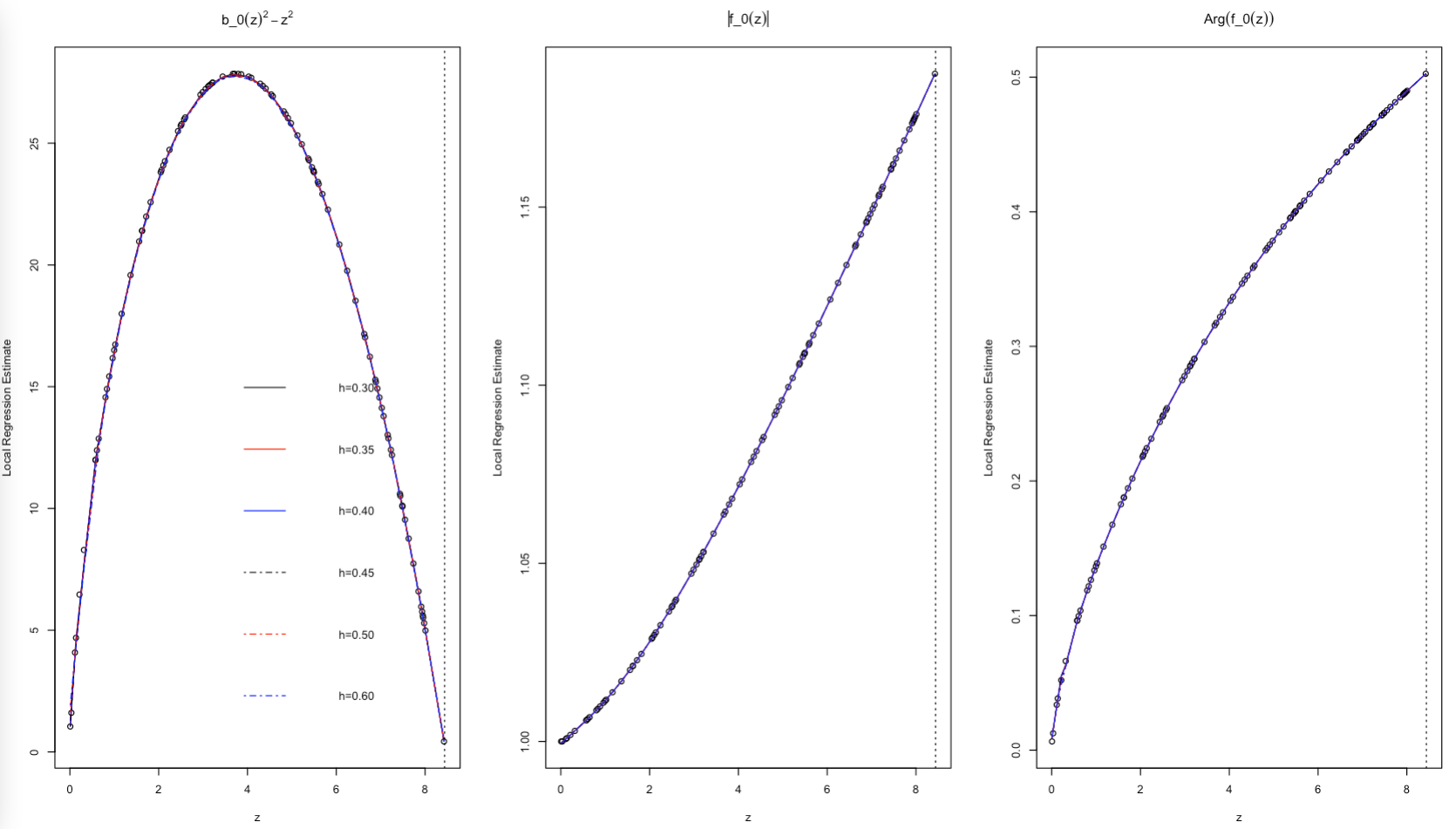}
\caption{\footnotesize{The estimates of critical collapse functions corresponding to $\gamma$ solution of hyperbolic case based on local 
regression method of order $l=1$ with bandwidth parameters $h=\{0.30,0.35,0.40,0.45,0.50,0.60\}$ 
based on a training sample of size $n=100$.}}
 \label{hy_ga_local}
\end{figure}

 Tables \ref{b_local}-\ref{argf_kernel} show the results of the numerical studies for all $l=1,\ldots,10$ and the top ten $h$ values. 
It is at once apparent that all the proposed estimators (excluding the polynomial of order $l=1$) perform 
very well in predicting the critical collapse functions in all elliptic and hyperbolic domains. The $\sqrt{\text{MSE}}$s of 
estimators are very small such that the polynomial regression, kernel regression, and local regression estimators can be 
considered almost unbiased in estimation of critical collapse functions even in the neighbourhood of the critical singularities. 
For graphical comparison of the proposed methods in estimating the critical functions, we presented the performance of the estimates in  Figures \ref{hy_al_local}-\ref{hy_ga_local} for each combination of the statistical methods, critical collapse functions and spaces. 
For example, Figure \ref{hy_al_local} shows the performance the local regression model in estimating the critical collapse functions.
The best performance of the local estimate appeared when $h$ was between $(0.07, 0.08)$; however, we intentionally selected   more widely spaced $h$ vales, namely ${0.07, 0.10, 0.25, 0.5, 0.75, 1.0}$  so that the human eyes can visually distinguish the curves. 
From Figure \ref{hy_al_local}, it is clear the $h$ values greater than $0.10$ result in over-smoothed estimates and consequently the prediction error increases. 
From Figures \ref{elip_poly}, \ref{elip_kernel} and \ref{elip_local}, one can compare graphically the performance of polynomial, kernel and local regression models in estimating the critical functions in elliptic space. 
From Figures \ref{hy_al_local}, \ref{hy_al_poly} and \ref{hy_al_kernel}, one can compare graphically the performance of the proposed models in estimating the critical functions corresponding to $\alpha$-solution of the hyperbolic space. 
From Figures \ref{hy_be_poly}, \ref{hy_be_kernel} and \ref{hy_be_local}, we can compare graphically the performance of the proposed statistical models in estimating the critical functions corresponding to the $\beta$-solution of the hyperbolic space. 
From Figures \ref{hy_ga_poly}, \ref{hy_ga_kernel} and \ref{hy_ga_local}, we can finally compare graphically the performance of the statistical models in estimating the critical functions corresponding to the $\gamma$-solution of the hyperbolic space.

The local regression estimators outperform the kernel and polynomial counterparts in estimation almost all three critical 
collapse functions in both elliptic and hyperbolic domains.  This superiority relies on the fact that the local regression 
estimator takes advantage of polynomial and kernel regression methods in estimation. 

While local and kernel regression methods estimate more accurately the critical collapse functions than polynomial 
regression method, polynomial regression method proposes closed form (and continuously differentiable) estimates 
for the critical functions. This closed and differentiable forms are of high importance  
 to make the critical solutions, critical exponents and the mass of Black Holes more tractable.

 The closed form polynomial regression estimates of order $l=6$ for critical collapse functions in elliptic domain are given by
\begin{align}\label{bhat_poly_el}
\widehat{b_0(z)^2} = 0.9791 + 0.4049 z + 8.5099 z^2 -10.6626 z^3 + 6.2498 z^4 -1.8050  z^5 + 0.2057 z^6.
\end{align}
\begin{align}\label{absfhat_poly_el}
\widehat{|f_0(z)|} = 0.9010 -0.2098 z -1.0300 z^2 + 1.6785 z^3 -1.1133 z^4 + 0.3462 z^5 -0.0414 z^6.
\end{align}
\begin{align}\label{argfhat_poly_el}
\widehat{\arg f_0(z)} = 0.0037 -0.1021 z + 0.3230 z^2 -0.0578 z^3 -0.0719 z^4 + 0.0398 z^5 -0.0060 z^6.
\end{align}
The closed form polynomial regression estimates of order $l=6$ for critical collapse functions corresponding to $\alpha$-solution domain in hyperbolic space are given by
\begin{align}\label{bhat_poly_hy_al}
\widehat{b_0(z)^2} =  1.0004 -0.0138 z + 0.5787 z^2 -0.3937 z^3 + 0.7078 z^4 -0.4632 z^5 + 0.1001 z^6.
\end{align}
\begin{align}\label{absfhat_poly_hy_al}
\widehat{|f_0(z)|} = 1.2258 -0.0136 z -0.1681 z^2 -0.3495 z^3 + 0.6597 z^4 -0.2450 z^5 -0.0025 z^6.
\end{align}
\begin{align}\label{argfhat_poly_hy_al}
\widehat{\arg f_0(z)} = 0.6167 -0.0157 z + 0.5834 z^2 -0.5672 z^3 + 1.1055 z^4 -0.9240 z^5 + 0.2412 z^6.
\end{align}
The closed form polynomial regression estimates of order $l=6$ for critical collapse functions corresponding to $\beta$-solution domain in hyperbolic space are given by
\begin{align}\label{bhat_poly_hy_be}
\widehat{b_0(z)^2} =  0.9051 + 1.9207 z + 5.6921 z^2 - 6.3768 z^3 + 3.1551 z^4 -0.7434 z^5 + 0.0678 z^6.
\end{align}
\begin{align}\label{absfhat_poly_hy_be}
\widehat{|f_0(z)|} = 1.0055 -0.0828 z + 0.3476 z^2 -0.2714 z^3 + 0.1069 z^4 -0.0213 z^5 + 0.0017 z^6.
\end{align}
\begin{align}\label{argfhat_poly_hy_be}
\widehat{\arg f_0(z)} = 0.0672 + 0.3845 z + 0.3860 z^2 -0.6077 z^3 + 0.3346 z^4 -0.0830 z^5 + 0.0078 z^6.
\end{align}
And eventually, the closed form polynomial regression estimates of order $l=6$ for critical collapse functions corresponding to $\gamma$-solution domain in hyperbolic space are given by
\begin{align}\label{bhat_poly_hy_ga}
\widehat{b_0(z)^2} =      1.9715 + 20.9801 z  -7.2930 z^2 +  2.2654 z^3 -0.3955 z^4 + 0.0350 z^5 -0.0012 z^6.
\end{align}
\begin{align}\label{absfhat_poly_hy_ga}
\widehat{|f_0(z)|} = 0.9997 + 0.0092  z + 0.0024 z^2 + 0.0001 z^3 -0.0001  z^4.
\end{align}
\begin{align}\label{argfhat_poly_hy_ga}
\widehat{\arg f_0(z)} =  0.0237 + 0.1287z -0.0204z^2 +  0.0023z^3 -0.0001z^4.
\end{align}

%%%%%%%%%%%%%%%%%%%%%%%%%%%%%%%%%%%%%%%%%%%%%%%%%%%%%%%%%%%%%%%%%%%%%%%%%%%%%%%%%%%%%
\section{Conclusion}\label{sec:sum}

The black hole solutions of axion-dilaton the system were recently investigated in elliptic and 
hyperbolic cases in four and five dimensions \cite{ours}. It is crucial for researchers to estimate 
the functional form of the critical collapse functions. These estimates pave the path to make the 
critical solutions, critical exponents, the mass of Black Holes and universality of Choptuik exponents
 more tractable. To our best knowledge, no research article in the literature investigated the properties 
 of nonlinear statistical models in estimating the critical collapse functions in Einstein-axion-dilaton.

 In this paper, we employed parametric polynomial regression, non-parametric kernel regression and semi-parametric local polynomial regression for the first time to estimate the functional forms of the critical collapse functions. 
From numerical studies, we observe that the local regression estimators outperform the kernel and polynomial counterparts in estimating almost all critical collapse functions in elliptic and hyperbolic domains. 
While local and kernel methods estimate more accurately the critical collapse function, the polynomial regression method enables us to obtain the closed-form and continuously differentiable estimates 
for the critical functions.
Given the closed forms of critical functions, a pressing question is if one can algebraically derive the critical exponents for the axion-dilaton system. 
Note that these methods are applied not only for Einstein-axion-dilaton system and similar solutions but also for other potential systems. These methods are  generic and can be used to any matter content for any space-time dimensions. This is a path that we plan to follow in the near future.

%%%%%%%%%%%%%%%%%%%%%%%%%%%%%%%%%%%%%%%%%%%%%%%%%%%%%%%%%%%%%%%%%%%%%%%%%%%%%%%%%%%%%
\section*{Acknowledgment}
Ehsan Hatefi would like to thank E. Hirschmann, R. Antonelli,  L. Alvarez-Gaume, and A. Sagnotti and R.
J. Lopez-Sastre for various useful discussions. Ehsan Hatefi would like to thank the International Maria Zambrano research grant and Armin Hatefi acknowledges the research support of the Natural Sciences and Engineering Research Council of Canada (NSERC).
  
%%%%%%%%%%%%%%%%%%%%%%%%%%%%%%%%%%%%%%%%%%%%%%%%%%%%%%%%%%%%%%%%%%%%%%%%%%%%%%%%%%%%%

%%%%%%%%%%%%%%%%%%%%%%%%%%%%%%%%%%%%%%%%%%%%%%%%%%%%%%%%%%%%%%%%%%%%%%%%%%%%%%%%%%%%%
\newpage

% latex table generated in R 4.1.1 by xtable 1.8-4 package
% Tue Oct 12 11:10:00 2021
\begin{table}[ht]
\centering
\footnotesize{\begin{tabular}{lccccc}
  \hline
  $l$   &   Elliptic &            & Hyperbolic &  \\ 
  \cline{3-5} 
        &            & $\alpha$-solution   & $\beta$-solution    & $\gamma$-solution \\ 
\hline
 1 & 0.7063268 & 0.0819795 & 1.0307328 & 6.3570261 \\ 
 2 & 0.0559801 & 0.0029784 & 0.0726698 & 0.5711519 \\ 
 3 & 0.0548808 & 0.0024525 & 0.0459405 & 0.2627902 \\ 
 4 & 0.0334353 & 0.0003811 & 0.0432772 & 0.1278611 \\ 
 5 & 0.0174070 & 0.0002913 & 0.0294283 & 0.0666275 \\ 
 6 & 0.0066110 & 0.0001207 & 0.0155586 & 0.0428675 \\ 
 7 & 0.0019199 & 0.0000196 & 0.0077121 & 0.0315803 \\ 
 8 & 0.0006227 & 0.0000030 & 0.0033251 & 0.0210668 \\ 
 9 & 0.0006231 & 0.0000026 & 0.0012466 & 0.0125086 \\ 
 10 & 0.0004316 & 0.0000018 & 0.0004576 & 0.0080795 \\ 
   \hline
\end{tabular}
\caption{\footnotesize{ The $\sqrt{\text{MSE}}$ of polynomial regression method of orders $l=1,\ldots,10$ in estimating 
the critical collapse response function $g(z)=b_0^2(z)-z^2$ in elliptic and hyperbolic cases based on a training sample of size $n=100$.}}
\label{b_poly}}
\end{table}

%%%%%%%%%%%%%%%%%%%%%%%%%%%%%%%%%%%%%%%%%%%%%%%%%%%%%%%%%%%%%%%%%%%%%%%%%%%%%%%%%%%%%
\newpage

% latex table generated in R 4.1.1 by xtable 1.8-4 package
% Tue Oct 12 11:10:00 2021
\begin{table}[ht]
\centering
\footnotesize{\begin{tabular}{lccccc}
  \hline
  $l$   &   Elliptic &            & Hyperbolic &  \\ 
  \cline{3-5} 
        &            & $\alpha$-solution   & $\beta$-solution    & $\gamma$-solution \\ 
\hline
1 & 0.0519709 & 0.0188249 & 0.0073093 & 0.0044686 \\ 
2 & 0.0118848 & 0.0092961 & 0.0051240 & 0.0012997 \\ 
3 & 0.0056050 & 0.0054771 & 0.0042886 & 0.0002376 \\ 
4  & 0.0053751 & 0.0020169 & 0.0021204 & 0.0000318 \\ 
5  & 0.0039364 & 0.0001069 & 0.0008176 & 0.0000250 \\ 
6  & 0.0020805 & 0.0001055 & 0.0002720 & 0.0000202 \\ 
7  & 0.0009439 & 0.0000180 & 0.0002590 & 0.0000148 \\ 
8  & 0.0002851 & 0.0000179 & 0.0002272 & 0.0000092 \\ 
9  & 0.0000928 & 0.0000055 & 0.0001656 & 0.0000047 \\ 
10 & 0.0000915 & 0.0000043 & 0.0000998 & 0.0000027 \\ 
  \hline
\end{tabular}
\caption{\footnotesize{The  $\sqrt{\text{MSE}}$ of polynomial regression method of orders $l=1,\ldots,10$ in estimating 
the critical collapse response function $g(z)=|f_0(z)|$ in elliptic and hyperbolic cases based on a training sample of size $n=100$.}}
\label{absf_poly}}
\end{table}

%%%%%%%%%%%%%%%%%%%%%%%%%%%%%%%%%%%%%%%%%%%%%%%%%%%%%%%%%%%%%%%%%%%%%%%%%%%%%%%%%%%%%
\newpage

% latex table generated in R 4.1.1 by xtable 1.8-4 package
% Tue Oct 12 11:10:00 2021
\begin{table}[ht]
\centering
\footnotesize{\begin{tabular}{lccccc}
  \hline
  $l$   &   Elliptic &            & Hyperbolic &  \\ 
  \cline{3-5} 
        &            & $\alpha$-solution   & $\beta$-solution    & $\gamma$-solution \\ 
\hline
1  & 0.0191634 & 0.0415803 & 0.0441231 & 0.0190365 \\ 
2  & 0.0124265 & 0.0200282 & 0.0151237 & 0.0046392 \\ 
3  & 0.0044581 & 0.0041933 & 0.0049122 & 0.0017751 \\ 
4  & 0.0011409 & 0.0011059 & 0.0043728 & 0.0008067 \\ 
5  & 0.0009282 & 0.0006399 & 0.0037591 & 0.0004130 \\ 
6  & 0.0007546 & 0.0000730 & 0.0024571 & 0.0002704 \\ 
7  & 0.0004682 & 0.0000358 & 0.0014501 & 0.0002036 \\ 
8  & 0.0001987 & 0.0000206 & 0.0007639 & 0.0001393 \\ 
9  & 0.0000671 & 0.0000197 & 0.0003697 & 0.0000851 \\ 
10  & 0.0000268 & 0.0000113 & 0.0001545 & 0.0000566 \\ 
  \hline
\end{tabular}
\caption{\footnotesize{The  $\sqrt{\text{MSE}}$ of polynomial regression method of orders $l=1,\ldots,10$ in estimating 
the critical collapse response function $g(z)=\arg f_0(z)$ in elliptic and hyperbolic cases based on a training sample of size $n=100$.}}
\label{argf_poly}}
\end{table}

%%%%%%%%%%%%%%%%%%%%%%%%%%%%%%%%%%%%%%%%%%%%%%%%%%%%%%%%%%%%%%%%%%%%%%%%%%%%%%%%%%%%%
\newpage
% latex table generated in R 4.1.1 by xtable 1.8-4 package
% Tue Oct 12 12:14:00 2021
\begin{table}[ht]
\centering
\footnotesize{\begin{tabular}{cccc}
\hline
   Elliptic &                  &Hyperbolic        &    \\ 
     \cline{2-4} 
            &$\alpha$-solution &$\beta$-solution  & $\gamma$-solution \\ 
\hline
 0.03150 (.06) & 0.00531  (.042) & 0.05777  (.110) & 0.45314  (.20) \\ 
 0.03291 (.07) & 0.00525  (.044) & 0.05781  (.112) & 0.46026  (.21) \\ 
 0.03414 (.08) & 0.00524  (.046) & 0.05783  (.114) & 0.44735  (.22) \\ 
 0.03359 (.09) & 0.00527  (.048) & 0.05775  (.116) & 0.43467  (.23) \\ 
 0.03413 (.10) & 0.00525  (.050) & 0.05767  (.118) & 0.43196  (.24) \\ 
 0.03585 (.11) & 0.00528  (.052) & 0.05761  (.120) & 0.43750  (.25) \\ 
 0.03823 (.12) & 0.00537  (.054) & 0.05764  (.122) & 0.45320  (.26) \\ 
 0.04059 (.13) & 0.00539  (.056) & 0.05777  (.124) & 0.47516  (.27)\\ 
 0.04272 (.14) & 0.00539  (.058) & 0.05775  (.126) & 0.49798  (.28) \\ 
 0.04479 (.15) & 0.00539  (.060) & 0.05767  (.128) & 0.51837  (.29) \\ 
 \hline
\end{tabular}
\caption{\footnotesize{The  $\sqrt{\text{MSE}}$ of kernel regression method evaluated at bandwidth $h$ (presented in parenthesis) in estimating 
the critical collapse response function $g(z)=b_0^2(z)-z^2$ in elliptic and hyperbolic cases based on a training sample of size $n=100$.}}
\label{b_kernel}}
\end{table}

%%%%%%%%%%%%%%%%%%%%%%%%%%%%%%%%%%%%%%%%%%%%%%%%%%%%%%%%%%%%%%%%%%%%%%%%%%%%%%%%%%%%%

\newpage
% latex table generated in R 4.1.1 by xtable 1.8-4 package
% Tue Oct 12 12:14:00 2021
\begin{table}[ht]
\centering
\footnotesize{\begin{tabular}{cccc}
\hline
   Elliptic &                  &Hyperbolic        &    \\ 
     \cline{2-4} 
            &$\alpha$-solution &$\beta$-solution  & $\gamma$-solution \\ 
\hline
 0.00324  (.06) & 0.00088  (.032) & 0.00227  (.07) & 0.00088  (.20) \\ 
 0.00346  (.07) & 0.00088  (.034) & 0.00230  (.08) & 0.00086  (.21) \\ 
 0.00339  (.08) & 0.00087  (.036) & 0.00240  (.09) & 0.00082  (.22) \\ 
 0.00342  (.09) & 0.00087  (.038) & 0.00248  (.10) & 0.00078  (.23) \\ 
 0.00350  (.10) & 0.00085  (.040) & 0.00250  (.11) & 0.00075  (.24) \\ 
 0.00380  (.11) & 0.00085  (.042) & 0.00249  (.12) & 0.00072  (.25) \\ 
 0.00411  (.12) & 0.00086  (.044) & 0.00250  (.13) & 0.00070  (.26) \\ 
 0.00437  (.13) & 0.00087  (.046) & 0.00249  (.14) & 0.00070  (.27) \\ 
 0.00452  (.14) & 0.00090  (.048) & 0.00248  (.15) & 0.00070  (.28) \\ 
 0.00471  (.15) & 0.00092  (.050) & 0.00255  (.16) & 0.00072  (.29) \\ 
 \hline
\end{tabular}
\caption{\footnotesize{The  $\sqrt{\text{MSE}}$ of kernel regression method evaluated at bandwidth $h$ (presented in parenthesis) in estimating 
the critical collapse response function $g(z)=|f_0(z)|$ in elliptic and hyperbolic cases based on a training sample of size $n=100$.}}
\label{absf_kernel}}
\end{table}

%%%%%%%%%%%%%%%%%%%%%%%%%%%%%%%%%%%%%%%%%%%%%%%%%%%%%%%%%%%%%%%%%%%%%%%%%%%%%%%%%%%%%

\newpage
% latex table generated in R 4.1.1 by xtable 1.8-4 package
% Tue Oct 12 12:14:00 2021
\begin{table}[ht]
\centering
\footnotesize{\begin{tabular}{cccc}
\hline
   Elliptic &                  &Hyperbolic        &    \\ 
     \cline{2-4} 
            &$\alpha$-solution &$\beta$-solution  & $\gamma$-solution \\ 
\hline
 0.00273  (.06) & 0.00379  (.024) & 0.00742  (.10) & 0.00397  (.20) \\ 
 0.00293  (.07) & 0.00379  (.026) & 0.00725  (.11) & 0.00403  (.21) \\ 
 0.00307  (.08) & 0.00350  (.028) & 0.00709  (.12) & 0.00392  (.22) \\ 
 0.00310  (.09) & 0.00352  (.030) & 0.00705  (.13) & 0.00378  (.23)\\ 
 0.00309  (.10) & 0.00356  (.032) & 0.00699  (.14) & 0.00373  (.24) \\ 
 0.00316  (.11) & 0.00366  (.034) & 0.00688  (.15) & 0.00376  (.25) \\ 
 0.00323  (.12) & 0.00374  (.036) & 0.00697  (.16) & 0.00387  (.26) \\ 
 0.00333  (.13) & 0.00379  (.038) & 0.00714  (.17) & 0.00405  (.27) \\ 
 0.00344  (.14) & 0.00373  (.040) & 0.00745  (.18) & 0.00424  (.28)\\ 
 0.00356  (.15) & 0.00364  (.042) & 0.00801  (.19) & 0.00439  (.29)\\ 
 \hline
\end{tabular}
\caption{\footnotesize{The  $\sqrt{\text{MSE}}$ of kernel regression method evaluated at bandwidth $h$ (presented in parenthesis) in estimating 
the critical collapse response function $g(z)=\arg f_0(z)$ in elliptic and hyperbolic cases based on a training sample of size $n=100$.}}
\label{argf_kernel}}
\end{table}

\newpage

\newpage

\newpage

%\newpage
%\begin{figure}
%\includegraphics[width=1.2\textwidth,center]{hyper_alpha_poly.png}
%\caption{\footnotesize{The estimates of critical collapse functions corresponding to $\alpha$ solution of hyperbolic case based on polynomial 
%regression method of orders $l=\{1,\ldots,6\}$ based on a training sample of size $n=100$.}}
% \label{hy_al_poly}
%\end{figure}

\newpage

%\newpage
%\begin{figure}
%\includegraphics[width=1.2\textwidth,center]{hyper_alpha_local.png}
%\caption{\footnotesize{The estimates of critical collapse functions corresponding to $\alpha$ solution of hyperbolic case based on local 
%regression method of order $l=1$ with bandwidth parameters $h=\{0.07,0.08,0.09,0.10,0.11,0.12\}$ 
%based on a training sample of size $n=100$.}}
% \label{hy_al_local}
%\end{figure}

\newpage

\newpage

\newpage

\newpage

\newpage

\newpage

\end{document}